\documentclass[twocolumn]{aastex62}
\pdfoutput=1

\usepackage [autostyle, english = american]{csquotes}
\MakeOuterQuote{"}

\received{January 5, 2019}

\accepted{April 12, 2019}

\submitjournal{ApJ}

\shorttitle{Probing shocks in Mrk 421}

\begin{document}



\title{Probing an X-ray flare pattern in Mrk 421 induced by multiple stationary shocks: a solution to the bulk Lorentz factor crisis}

\correspondingauthor{Olivier Hervet}
\email{ohervet@ucsc.edu}

\author{Olivier Hervet}
\affil{Santa Cruz Institute for Particle Physics and Department of Physics\\
 University of California\\
  Santa Cruz, CA 95064, USA}

\author{David A. Williams}
\affil{Santa Cruz Institute for Particle Physics and Department of Physics\\
 University of California\\
  Santa Cruz, CA 95064, USA}
  
\author{Abraham D. Falcone}
\affil{Department of Astronomy and Astrophysics\\
 Pennsylvania State University\\
University Park, PA 16802, USA}
  
\author{Amanpreet Kaur}
\affil{Department of Astronomy and Astrophysics\\
 Pennsylvania State University\\
University Park, PA 16802, USA}



\begin{abstract}

The common observations of multiple radio VLBI stationary knots in high-frequency-peaked BL Lacs (HBLs) can be interpreted as multiple recollimation shocks accelerating particles along jets. This approach can resolve the so-called "bulk Lorentz factor crisis" of sources with high Lorentz factor, deduced from maximum $\gamma-\gamma$ opacity and fast variability, and apparently inconsistent slow/stationary radio knots.
It also suggests that a unique pattern of the non-thermal emission variability should appear after each strong flare. 
Taking advantage of the 13 years of observation of the HBL Mrk 421 by the X-ray Telescope on the Neil Gehrels Swift Observatory (\textit{Swift}-XRT), we probe for such an intrinsic variability pattern.
Its significance is then statistically estimated via comparisons with numerous similar simulated lightcurves.
A suggested variability pattern is identified, consistent with a main flare emission zone located in the most upstream 15.3 GHz radio knot at $0.38$ mas from the core.
Subsequent flux excesses in the lightcurve are consistent with a perturbation crossing all the downstream radio knots with a constant apparent speed of $45$ c.
The significance of the observed variability pattern not arising from stochastic processes is found above 3 standard deviations, opening a promising path for further investigations in other blazars and with other energy bands.
In addition to highlight the role of stationary radio knots as high-energy particle accelerators in jets, the developed method allows estimates of the apparent speed and size of a jet perturbation without the need to directly observe any motion in jets.

\end{abstract}



\section{Introduction} \label{sec:intro}

Multiwavelength studies of the variability and modeling of radio-loud AGN broadband SEDs attest to a compact emission zone moving with a high Lorentz factor close to the central engine. 
The particle individual Lorentz factors are often estimated to be above $10^6$ for the most energetic blazars, implying long-standing and powerful particle acceleration mechanisms.
While the scenario of magnetic reconnection has received considerable attention during recent years, due to recent progress with MHD simulations \citep{Sironi_2014}, the scenario of acceleration by shocks remains the most studied and the most accepted for the typical activity state of radio-loud AGN and their common variability \citep{Marscher_1985, Spada_2001, Fromm_2011}.

The shock scenario is supported by multiple observations of gamma-ray flares in coincidence with the emergence of a jet perturbation (or overdensity) in or close to the radio core, mainly seen in flat-spectrum radio quasars (FSRQs) and some low- or intermediate-frequency-peaked BL Lacs (LBLs and IBLs) \citep{Jorstad_2001, Marscher_2008,Abeysekara_2018}.
The formation of recollimation shocks (also referenced as conical standing shock or reconfinement shock) in jets is also a phenomenon naturally observed in hydrodynamic and magnetohydrodynamic jet simulations as soon as a supersonic, or super-Alfvenic, non-pressured matched flow propagates through an external medium.
This pressure mismatch at the interface between the jet inlet and the external medium generates two conical waves, namely a shock wave and a rarefaction wave. The shock wave propagates toward the external medium, and is reflected toward the jet axis as it reaches equilibrium with the external medium pressure. The rarefaction wave propagates
toward the jet axis, locally dropping the jet pressure and accelerating the flow. The flow is then significantly slowed down after it reaches the reflection point of the conical waves at the jet axis. This process repeats and can produce a string of recollimation shocks until the full dissipation of energy carried out by the waves \citep[e.g.][]{Falle_1991, vanPutten_1996, Gomez_1997, Mizuno_2015, Hervet_2017}.

Contrary to other blazar types, high-frequency-peaked BL Lacs (HBLs) show mainly stationary or low-speed VLBI radio features (radio knots) in their jets, in stark contrast to the high Lorentz factor values deduced from their variability or SED modeling \citep{Hervet_2016, Piner_2018}. 
Most of the interpretations of this issue imply two distinct regions between radio knots and high-energy emission zones. Slow/stationary radio knots are assumed to come from a slower and wider jet part than the high energy emission zone. It can be understood as a strong jet deceleration very close to the core  \citep{Georganopoulos_2003}, or a stratified jet with differential speeds, as non-steady outflows \citep{Lyutikov_2010} or spine-layer structure \citep{Ghisellini_2005,Piner_2018}.
We adopt the interpretation of slow/stationary radio knots as a multiple recollimation shock structure, very stable for these sources due to their lower outer-jet kinetic power \citep{Hervet_2017}.

Following the shock-in-jet model developed by \cite{Marscher_1985}, a flare should happen when a perturbation (or moving shock) passes trough a recollimation shock. This scenario was adapted and improved by many further works and is quite successful as a picture of the the general broadband blazar flaring behavior \citep[e.g.][]{Komissarov_1997,Turler_2000, Turler_2011, Nalewajko_2009, Nalewajko_2012, Fromm_2011, Fromm_2016, Marscher_2014}. Successive flares are then believed to be triggered by a stochastic injection from the central engine.
However, while this approach assumes one shock at the base of the jet is responsible for the main dissipation process, it does not consider the other potential flares produced by downstream shocks. 
We investigate here the possibility of successive flares associated with successive recollimation shocks in relativistic jets.
If we relate stationary radio knots to recollimation shocks, we can predict a distinct pattern of variability based on inter-knot gaps.
Thus, after each strong flare occurring at the base of the jet, one should detect several other flares in accordance with the VLBI radio knot distribution in the jet, for a given velocity of the flow.
The confirmation of such a pattern in HBL lightcurves would validate the role of stationary radio knots as high-energy particle accelerators, and characterize the apparent speed and size of underlying perturbations, extremely valuable for constraining the modeling parameters.

In Section \ref{Section::Method} we introduce the basic concept of the proposed scenario and the ideal source for its application, Mrk 421. 
In Section \ref{Section::Data} and \ref{Section::Formatting_the_dataset} we describe how X-ray long-term lightcurves are handled in view of having the most efficient probe to detect a possible intrinsic post-flare variability pattern. The theoretical models used to check our scenario are developed in Section \ref{Section::model}.
In Section \ref{Section::Simulation} we describe the method used to create simulated lightcurves as similar as possible to the real dataset, and also discuss biases induced by these simulations.
Results and a general discussion are in Section \ref{Section::Discussion}.

Throughout this paper, a flat $\mathrm{\Lambda C D M}$ cosmology is adopted with $\mathrm{H_0} = 69.7$ km s$^{-1}$ Mpc$^{-1}$, $\Omega_M = 0.286$, and $\Omega_v = 0.714$ \citep{Bennett_2014}. It leads to a projected scale of 1 mas = 0.603 pc at the redshift $z=0.030$ of Mrk 421.

\section{Method and application to Mrk 421}
\label{Section::Method}

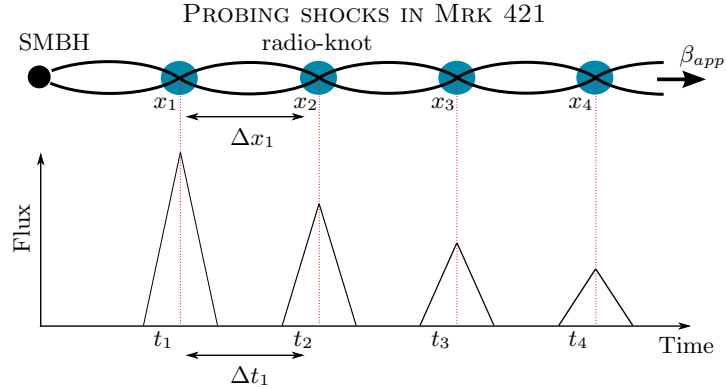
\begin{figure*}[t!]
\begin{center}
	\includegraphics[width= 0.5\textwidth]{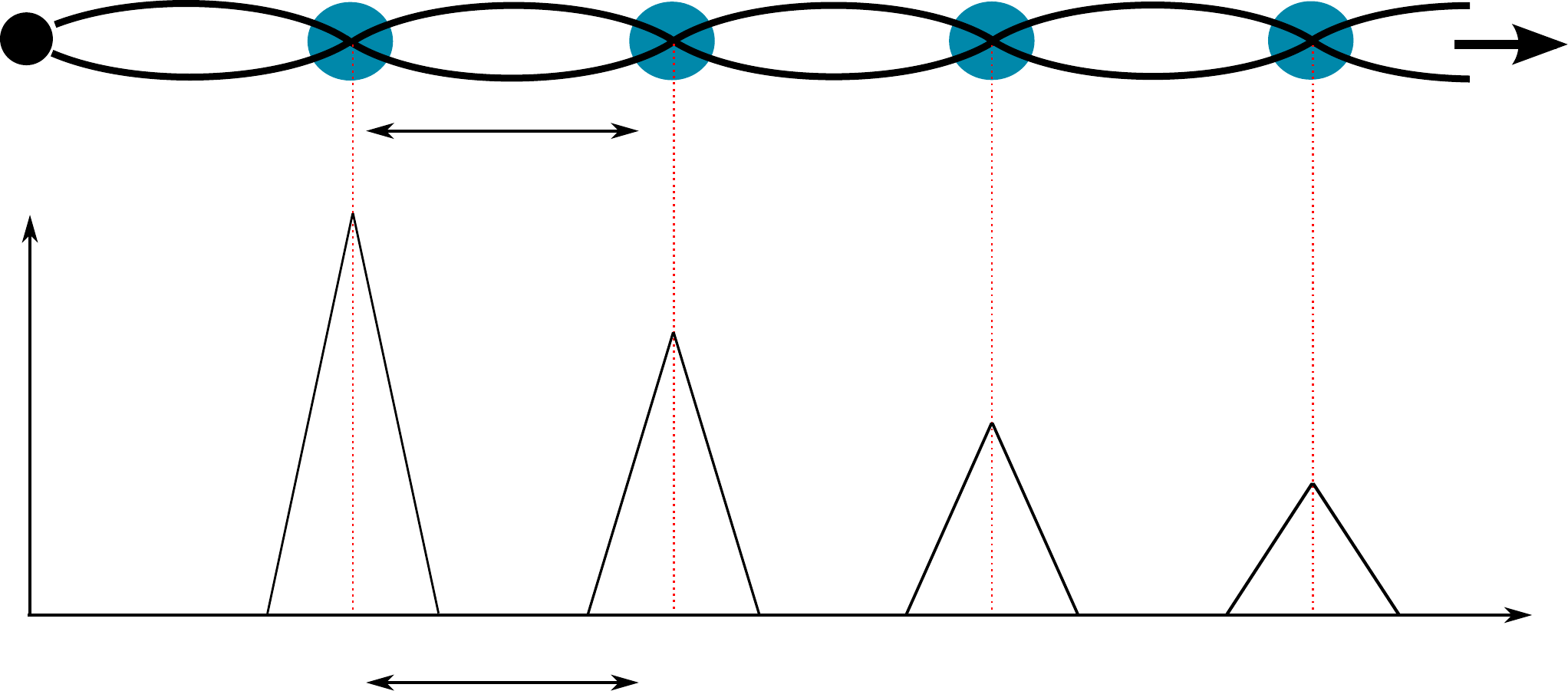}
			\put(-260,117){\makebox(0,0)[lb]{SMBH}}
			\put(-168,117){\makebox(0,0)[lb]{radio-knot}}
			\put(-10,110){\makebox(0,0)[lb]{$\beta_{app}$}}
	 		\put(-209,94){\makebox(0,0)[lb]{$x_1$}}
	 		\put(-156,94){\makebox(0,0)[lb]{$x_2$}}
	 		\put(-104,94){\makebox(0,0)[lb]{$x_3$}}
	 		\put(-52,94){\makebox(0,0)[lb]{$x_4$}}
	 		\put(-180,80){\makebox(0,0)[lb]{$\Delta x_1$}}
	 		\put(-209,4){\makebox(0,0)[lb]{$t_1$}}
	 		\put(-156,4){\makebox(0,0)[lb]{$t_2$}}
	 		\put(-104,4){\makebox(0,0)[lb]{$t_3$}}
	 		\put(-52,4){\makebox(0,0)[lb]{$t_4$}}
	 		\put(-180,-10){\makebox(0,0)[lb]{$\Delta t_1$}}
	 		\put(-262,40){\makebox(0,0)[lb]{\rotatebox{90}{Flux}}}
	 		\put(-18,2){\makebox(0,0)[lb]{Time}}
 	\caption{Simplified scheme of the expected lightcurve signature of a perturbation crossing the knots $x_i$ with an apparent speed $\beta_{app}$ linking the inter-knot distance $\Delta x_1$ with the delay between two consecutive flares $\Delta t_1$.}
 		\label{Fig::schema_methode}
 \end{center}
\end{figure*}

\subsection{Concept of the method}

The core of the method is to probe flares associated with the flow passing through the knots, assuming they are stationary shocks. For a given apparent speed $\beta_{app}$, the time delay of the secondary flares can be set knowing the radio knot positions, as shown in Figure \ref{Fig::schema_methode}.

Considering a constant speed of the flow through a straight jet, the time gap $\Delta t$ between each successive flare in the lightcurve should be directly proportional to the observed inter-knot gap $\Delta x$. We have the relation

\begin{equation}
\label{Eq::TimeGap}
\Delta t_i = (1 + z) \frac{\Delta x_i}{c \beta_{app}}.
\end{equation} 

Considering the association of radio knots with recollimation shocks, the underlying flow is expected to accelerate upstream of each shock due to the presence of rarefaction waves locally decreasing the pressure. The speed should then decrease after the shock. The realistic speed profile would be an oscillation, likely with a slower acceleration due to global conical opening of the jet \citep{Komissarov_1997, Gomez_1997, Mizuno_2015, Hervet_2017}.
Throughout this paper we consider the approximation of an average constant speed of the underlying flow valid, with the main motivation keeping the lightcurve model developed in Section \ref{Section::model} as simple as possible.
This approximation can be supported with the observed motions in radio jets, which in the majority are well fitted by a constant-speed motion \citep{Lister_2016}.
As further discussed in Section \ref{Section::model}, the theoretical model developed also considers the width of the peaks from the size of the radio knots and a damping factor between successive flares.

\subsection{Mrk 421: the ideal candidate}

\begin{figure*}
\begin{center}
	\begin{minipage}[b]{0.49\linewidth}
   		\centering \includegraphics[width=4.7cm]{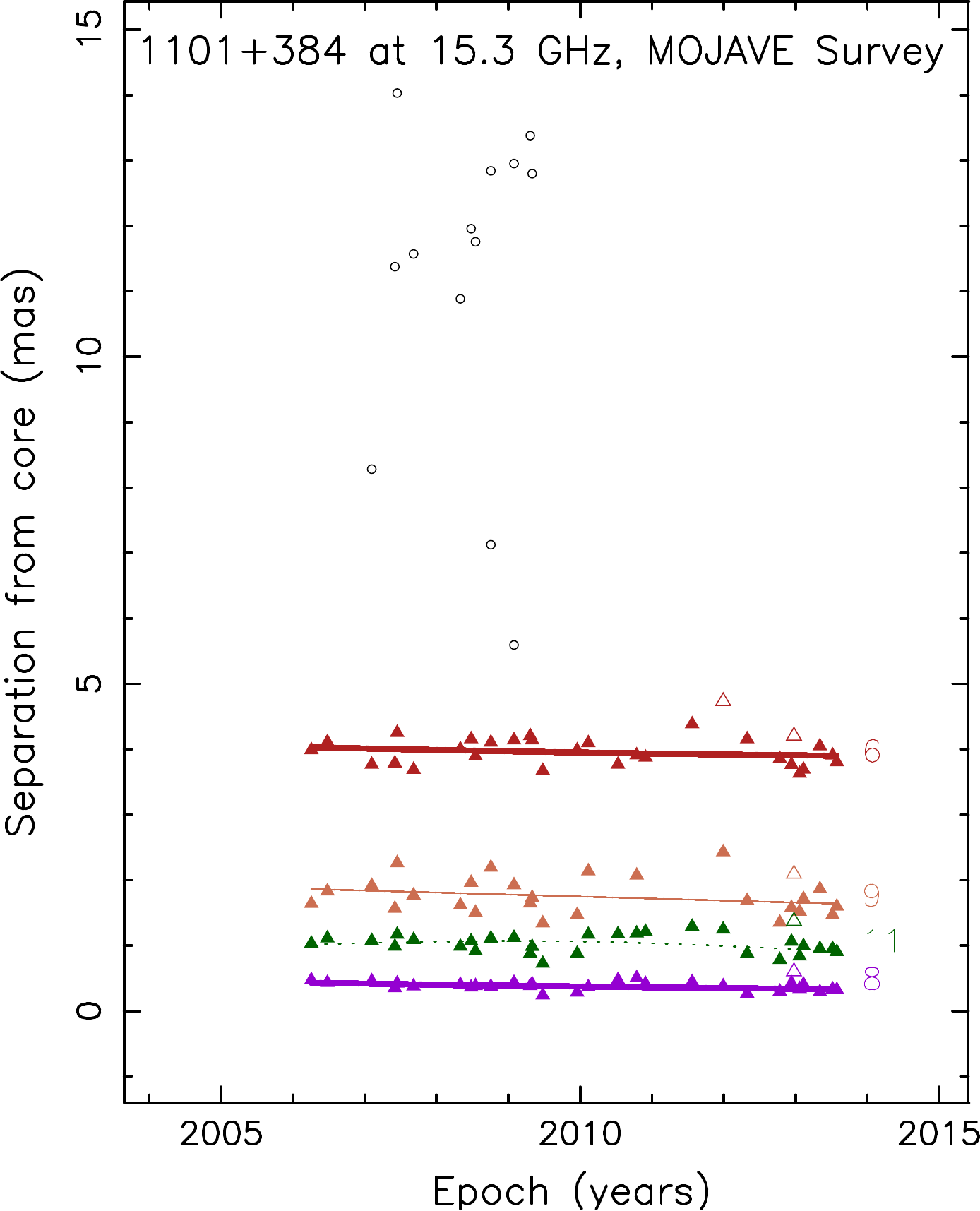}
	\end{minipage}\hfill
	\begin{minipage}[b]{0.49\linewidth}
      \centering \includegraphics[width=8cm]{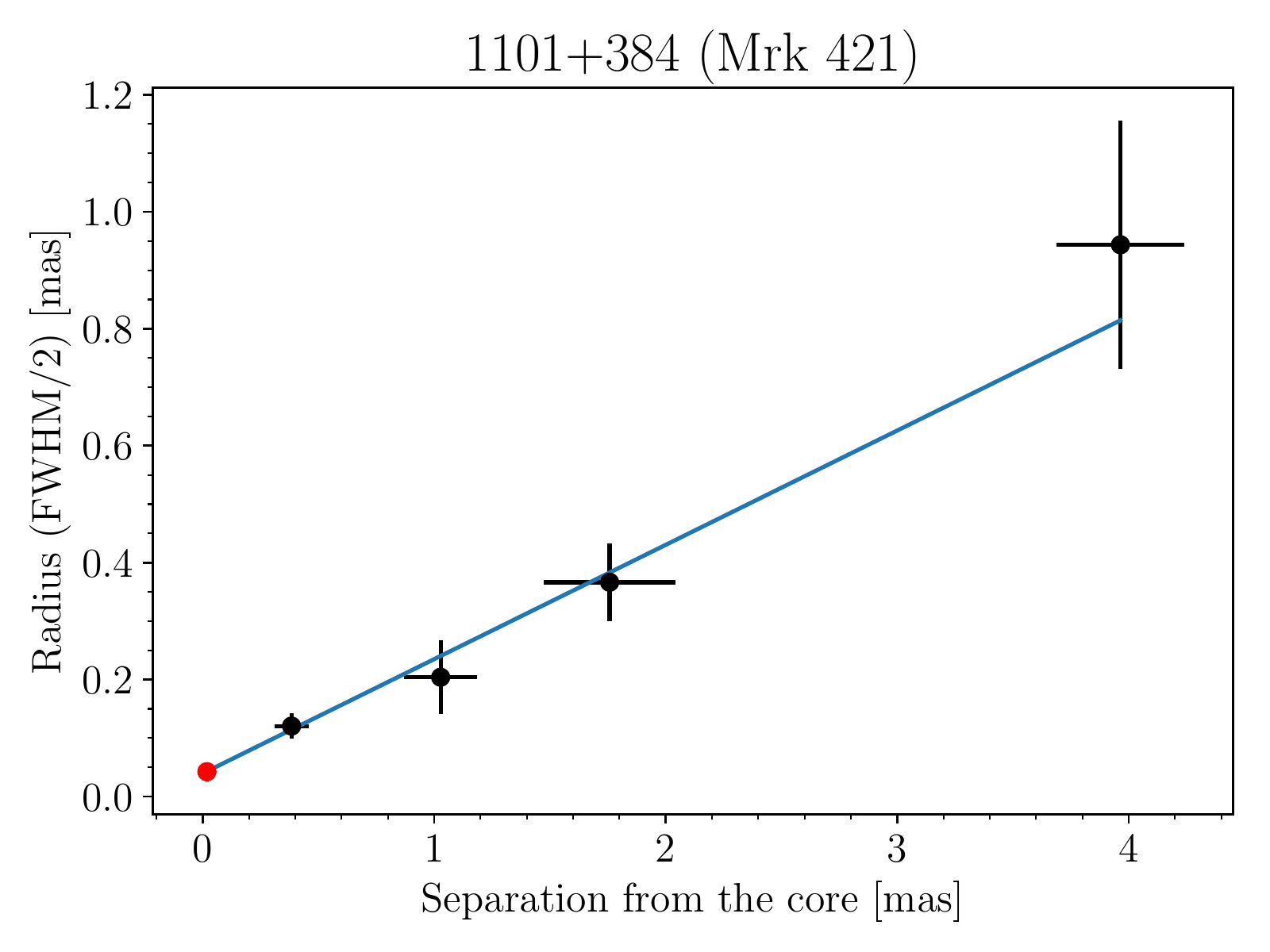}
	\end{minipage}\hfill
	\hspace{0.5cm}
 		\caption{\textit{Left:} Temporal evolution of knot-core distances of Mrk 421 observed by MOJAVE. Four quasi-stationary knots are firmly detected within 5 mas of the radio core between 2006 and 2014. Grey dots are considered as non-robust features to measure the jet kinematics.
 		\textit{Right:} Mean core distance and radius of radio knots with standard deviation, fitted by a linear function. The red point is the radio core. Adapted from \cite{Lister_2016}.}
 		\label{Fig::knots_Mrk421}
 \end{center}
\end{figure*}

Mrk 421 is the brightest X-ray and gamma-ray HBL in the sky in its flaring and average state \citep{Stroh_2013}. It is one of the most monitored blazars in all wavelengths and shows frequent giant flares \citep[e.g.][]{Aleksic_2015, Abeysekara_2017,Fraija_2017}.
Mrk 421 is perfectly adapted for this study by also presenting 4 well-defined VLBI quasi-stationary knots within 5 mas of the radio core at 15.3 GHz, as shown in Figure \ref{Fig::knots_Mrk421} (\textit{Left}) from the MOJAVE collaboration.\footnote{\label{note1}\url{http://www.physics.purdue.edu/MOJAVE}}
All the observed knots show either non-radial or downward motions. Such motions would be very challenging to be described with a ballistic model, but can naturally match low amplitude shifts/oscillations of quasi-stationary recollimation shocks. The fastest measured knot measured in VLBI ($\#$6) displays an apparent speed of $0.217 \pm 0.026$ c, roughly perpendicular to the jet direction \citep{Lister_2016}, and the usual Doppler factor deduced from broadband spectral energy distribution (SED) modeling is about 20-25 \citep{Blazejowski_2005, Balokovic_2016, Carnerero_2017,Kapanadze_2018_2,Kapanadze_2018_1}, 
which can be seen as a lower limit, since the Doppler factor is usually constrained from the shortest variability timescale observed and from the maximum possible photon-photon opacity within the emitting region.
For a canonical blazar angle with the line of sight of 2 deg, the SED models lead to a Lorentz factor $\Gamma_{model} \gtrsim 14$, which should be related to apparent downstream speed of $\beta_{app} \gtrsim 11$ c.  Mrk 421 is then strongly affected by the bulk Lorentz factor crisis, which is ideal for our study.

For this study we consider these 4 knots as stationary recollimation shocks with their distance to the radio core given by the mean value of the measured distances from the MOJAVE Collaboration. The uncertainty on their distance to the core and radius are given by the standard deviation of the dataset. The Mrk 421 knot string follows a conical expansion well, as shown in Figure \ref{Fig::knots_Mrk421} (\textit{Right}). The knots' radius is fitted by a linear function  $f(x) = (0.195 \pm 0.015) x + (3.94\pm 0.76) \times 10^{-2}$ mas, with a reduced $\chi^2$ of 0.28. 
The radio knot positions of Mrk~421 were measured in several other studies for different frequencies and epochs. Although the MOJAVE dataset is the one the most simultaneous with the lightcurve in our study, it remains relevant to check the consistency of these measurements with the previous observations described in \cite{Piner_2010} (with extended dataset from \cite{Piner_1999, Piner_2005}), and  \cite{Lico_2012}.

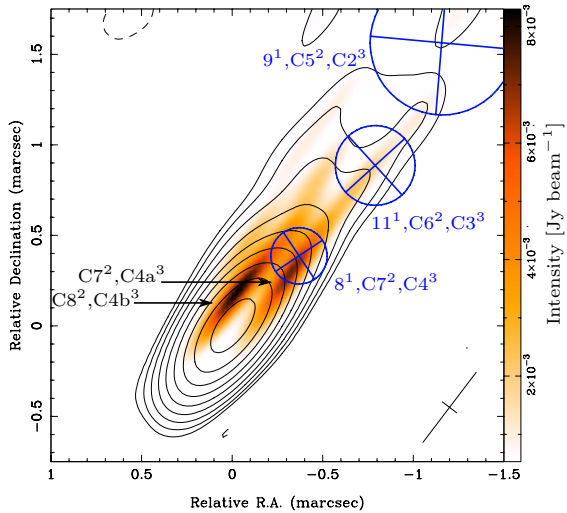
\begin{figure}[h!]
\begin{center}
	\includegraphics[width= 7cm]{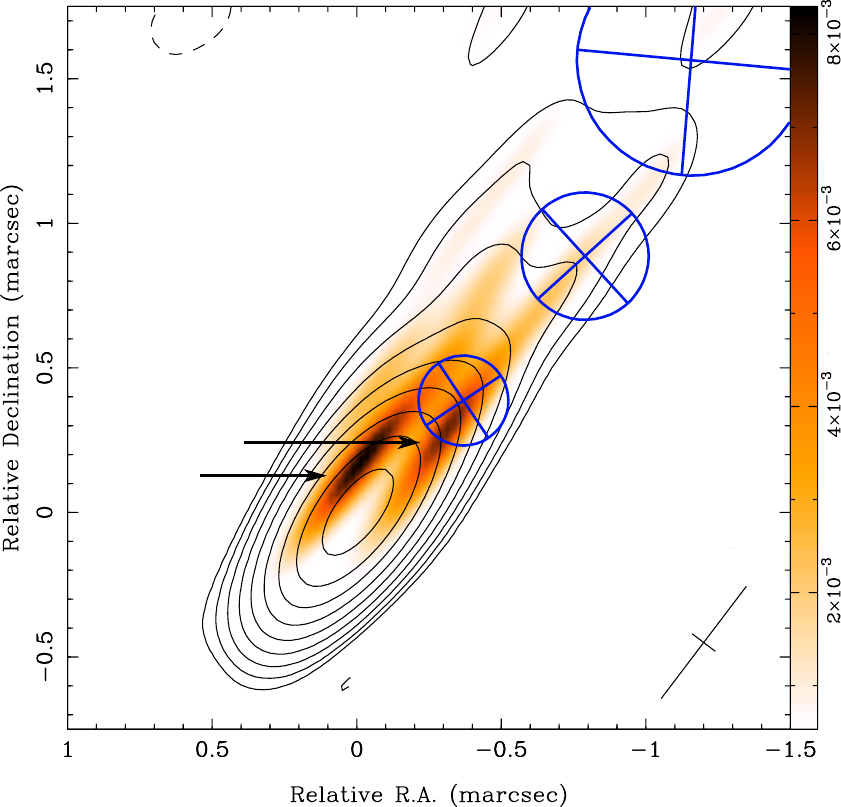}
	\put(-183,76){\makebox(0,0)[lb]{\scriptsize{C8$^2$,C4b$^3$}}}
	\put(-173,85){\makebox(0,0)[lb]{\scriptsize C7$^2$,C4a$^3$}}
	\put(-76,82){\makebox(0,0)[lb]{\scriptsize  \textcolor{blue}{8$^1$,C7$^2$,C4$^3$}}}
	\put(-62,105){\makebox(0,0)[lb]{\scriptsize  \textcolor{blue}{11$^1$,C6$^2$,C3$^3$}}}
	\put(-103,167){\makebox(0,0)[lb]{\scriptsize  \textcolor{blue}{9$^1$,C5$^2$,C2$^3$}}}
	\put(3,70){\makebox(0,0)[lb]{\rotatebox{90}{\scriptsize  Intensity [Jy beam$^{-1}$]}}}
 	\caption{43 GHz radio map contour with color image representing the intensity after removing the core emission, observed the 2008 Dec. 8 \citep{Piner_2010}. Blue circles are 15.3 GHz components fitted by 2D Gaussian, from 2011 Jan. 14 \citep{Lico_2012}. The various knot IDs follow the references: MOJAVE, 1; \cite{Piner_2010}, 2; \cite{Lico_2012}, 3. We can note that the 15.3 GHz knots presented here slightly differ in size and position from the 7-year-average values we are using in our study.}
 		\label{Fig::radio_map}
 \end{center}
\end{figure}

\begin{table}
\footnotesize
\renewcommand{\thetable}{\arabic{table}}
\centering
\caption{Projected distance from the radio core of the 4 VLBI quasi-stationary radio knots referenced by MOJAVE with their different associated names.} 
\label{Tab::knots_position}
\begin{tabular}{c|c|c|c|c}
\tablewidth{0pt}
\hline
\hline
knot \# & knot \# & knot \# & core distance & radius\\
(1) & (2) & (3)  & [mas] (1) & [mas] (1)\\
\hline
\decimals
Core  & - & - & - & $4.24 \pm 1.62 \times 10^{-2}$\\
\hline
8  & C7    & C4      & $0.38 \pm 0.07$ & $1.20 \pm 0.22 \times 10^{-1}$\\
11 & C6    & C3      & $1.03 \pm 0.16$ & $2.04 \pm 0.63 \times 10^{-1}$\\
9  & C5    & C2      & $1.76 \pm 0.29$ & $3.66 \pm 0.66 \times 10^{-1}$\\
6  &       & C1      & $3.96 \pm 0.28$ & $9.44 \pm 0.21 \times 10^{-1}$\\
\hline
\multicolumn{5}{l}{\footnotesize{1: MOJAVE, 2: \cite{Piner_2010}, 3: \cite{Lico_2012}}}
\end{tabular}
\end{table}

\cite{Piner_2010} reported VLBA observations at 22 GHz and 43 GHz of Mrk 421 between 1994 and 2009. They observed knots consistent with the ones detected by MOJAVE, they however detect a supplementary component between 2008-2009 at 43.2 GHz, C8,  at $\sim 0.2$ mas from the core.
\cite{Lico_2012}, who performed VLBA observations in 2011, have similar observations. While their 15.36 GHz analysis is consistent with the one presented by MOJAVE, at 23.804 GHz the first radio knot can be divided as 2 distinct components named C4a and C4b.
\cite{Piner_2010} noticed that these 43.2 GHz knots C7 and C8 (or C4a C4b from \cite{Lico_2012}) can be associated with the eastern and western limb-brightened jet structure of the jet (see Figure \ref{Fig::radio_map}).

\begin{figure*}[t!]
\begin{center}
	\includegraphics[width= \textwidth]{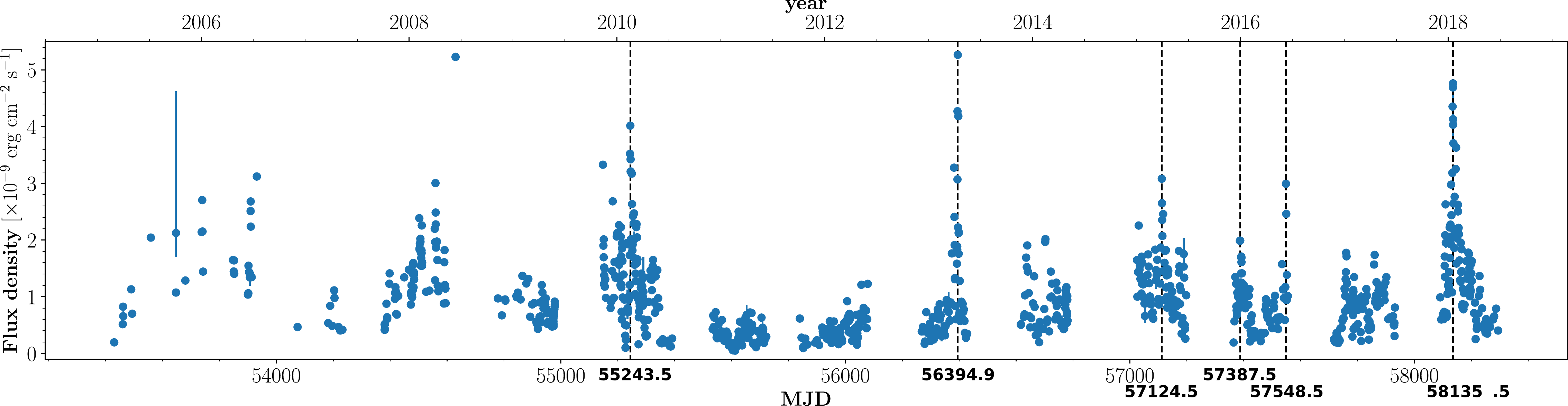}
 	\caption{\textit{Swift}-XRT lightcurve from March 2005 to May 2018. Vertical dashed lines represent the selected flares as discussed in Section \ref{Section:flare_selection}.}
 		\label{Fig::Lightcurve}
 \end{center}
\end{figure*}

The limb-brightened emission is likely an indication of a spine-sheath jet where the outer jet is either more Doppler boosted (due to a smaller angle with the line of sight), or presents a larger intrinsic synchrotron emissivity. Throughout this study, we consider this local limb-brightened emission at high frequencies as a single shock in the inner jet, associated with the position of the knot 8.
For more clarity we reference the studied knot positions given by MOJAVE in Table \ref{Tab::knots_position} with their associated names from previous studies.

The high-energy emission zone location(s) of radio-loud AGN is still an unresolved question. Multiple studies have highlighted the likely presence of multiple high-energy zones within the jets from broadband emission models and variability studies \citep[e.g.][]{Raiteri_2010,Tavecchio_2011, Nalewajko_2012, Hervet_2015}.
When comparing the high- and very high-energy flares with radio VLBI measurements, it appears that flares can be either associated with the radio core or a radio knot outside the core
\citep[e.g.][]{Abramowski_2012,Marscher_2014}. 
We note that the radio core is by definition ambiguous and can be itself composed of several radio knots when observed with better angular resolution \citep{Gomez_2016}.
Not knowing if the radio core of Mrk 421 could be associated with a strong first recollimation shock, we probe the two following hypotheses:
\begin{itemize}
\item[$\bullet$] The biggest observed flares are produced in the radio core; then 4 following flares are expected in the lightcurve.
\item[$\bullet$] The biggest observed flares are produced in the first radio knot; then 3 following flares are expected in the lightcurve.
\end{itemize}

\section{Swift-XRT analysis}
\label{Section::Data}

The X-ray Telescope on the Neil Gehrels Swift Observatory \citep{Burrows_2005} is sensitive in the soft X-ray energy range (0.3 -10 keV), which is excellent for measuring flux at, or near, the synchrotron peak energy for HBLs such as Mrk 421.  Large amplitude flares typically produce copious synchrotron emission in this energy band.  \textit{Swift}-XRT has proven to be highly capable at monitoring both long-term flux variability (with a baseline of $>14$ years) and large amplitude flares with precise flux and spectral measurements.\footnote{\url{https://www.swift.psu.edu/monitoring/}}

Since Mrk 421 is typically at a high enough count rate to induce pile-up of photons in photon counting mode, most of the observations were taken in Window Timing (WT) mode.
The cleaned level-3 event files were used for extracting data products. Initially, a cleaned event file was separated into individual snapshots (i.e. individual pointed observations). Each snapshot was then utilized to extract an image within the 0.3 - 10 keV energy range. The first 150 seconds of data were discarded from each
snapshot for the WT mode observations in order to exclude data with any spacecraft settling issue that might have occurred during this time interval. A pile-up correction was performed using the method described in \cite{Romano_2006}. The extracted spectrum for each snapshot was obtained by selecting a box with dimensions $40 \times 20$ pixels (2.36 arcsec/pixel). The source box region was rotated as per roll angle for the given snapshot. An annular boxed background region rotated at the same angle, with size 100 pixels (same height as source region; 20 pixels) was used to obtain the background spectrum. 

For observations taken in Photon Counting (PC) mode, first a circular source region with size 20 pixels and an annular background region were chosen to extract spectra. If the source counts were found to be $> 0.6$ cts/s, a pile-up correction was performed. In order to correct for pile-up, an appropriate annular region was selected as the source region for the final spectrum extraction, ensuring that the count rate drops to at least 0.6 counts/s. %

Fluxes and spectra are extracted with 1-day binning. XSpec \citep{Arnaud_1996} was utilized to fit all spectra with a model comprised of a log parabola combined with absorption as specified in the Tuebingen-Boulder ISM absorption model. This X-ray spectral shape of Mrk 421 is confirmed by previous studies \citep{Massaro_2004}. The hydrogen column density was fixed to 0.019 cm$^{-2}$ , which was derived from the LAB survey \citep{Kalberla_2005}. Within XSpec, we utilized cflux to determine the unabsorbed flux in the 0.3 - 10 keV band.

The full \textit{Swift}-XRT lightcurve is shown in Figure \ref{Fig::Lightcurve}.

\section{Formatting the dataset}
\label{Section::Formatting_the_dataset}

\begin{figure*}[t!]
\begin{center}
	\includegraphics[width= \textwidth]{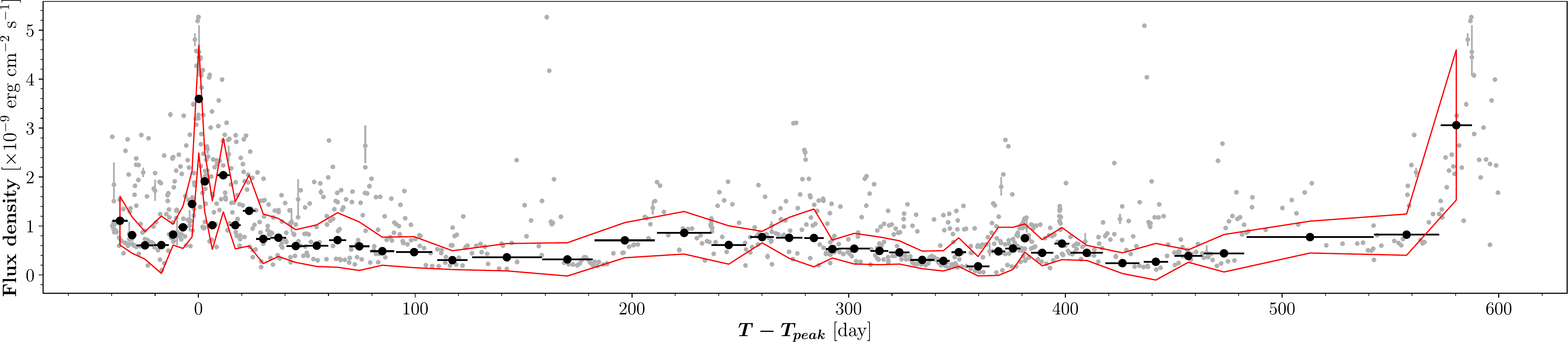}
 	\caption{Flare-stacked lightcurve used to probe a post-flare variability pattern.  For clarity we show a binned dataset, with 18 data points per bin. The red lines picture the RMS range associated with the flux dispersion of stacked lightcurves.}
 		\label{Fig::stacked_lightcurve}
 \end{center}
\end{figure*}

\subsection{Major flare selection}
\label{Section:flare_selection}

The brighter a flare is, the more we expect the flare to be associated with an ejection through the main recollimation shock. Therefore we want to select the brightest flares in X-rays as input to our method.
The way the flares are selected can impact the results of the study.
Selecting too few flares will not bring enough constraints on the method, with the risk of being biased against the typical behavior of the source by selecting "exceptional" events.
On the other hand, selecting many weak flares increases the risk of injecting intrinsic stochastic fluctuations into the method and burying any possible variability pattern in noise.

As a middle ground, we select a flare only if the peak of flux is above the 90th percentile of the distribution, giving a threshold value of $1.90 \times 10^{-9}$ erg cm$^{-2}$ s$^{-1}$ for the \textit{Swift}-XRT dataset. Later sections discuss the impact of using a different flux threshold to select flares. No flare is considered if it has a significance less than 3 sigma above the median flux.
Also, in order to have confidence that a high measured flux is the flare peak, a flare is selected only if it has at least 1 data point in the 10 days before it and 3 data points in the 10 days after it.
This ensures having a temporal estimation ($\sim$ day) of a flare, which is relevant for the lightcurve analysis method, as developed in the following section.

Finally, we intend to select the first flare which starts a sequence and to avoid having two series too close to each other, which can mislead the method.
It can be done by selecting only the strongest flare in a given time range.
Thus, a flare is not selected if it happens during the 100 days before or after a stronger flare. This exclusion zone of 100 days apply even if the stronger flare was not selected for our method (due to a bad timing estimation).
This cut, however, has some limitations. In case of low apparent speeds of the flow ($\lesssim 5$ c), the time-gap expected between the nearby radio knots of Mrk 421 is more than 100 days, making our method less sensitive for those speeds.
And having too big a time-gap would lead to a limited number of flares.
The impact of the choice of these cuts on the final results is quantified in Section \ref{Section::Systematics} to estimate systematic errors.

The date of selected flares and their associated position in the lightcurve are shown in Figure \ref{Fig::Lightcurve}.

\subsection{Lightcurve stacking}

This study aims to see if there is a regular intrinsic pattern in the lightcurve after a strong flare. 
However, each X-ray flare of Mrk 421 is different with a variability apparent afterward that we assume is in part due to strong and unpredictable turbulences within the jet. 
Also, many observing gaps after flares, which can be as big as several months, make the definition of a variability pattern even more difficult in a flare-by-flare study.

Hence we stack all the selected flares on the dates given in Figure \ref{Fig::Lightcurve}. By working on this stacked lightcurve, we expect that the pure stochastic variability will play a reduced role and that we have a typical post-flare dataset without large observing gaps. 
There is a risk that an uneven stacked dataset creates a misleading pattern, not associated to any physical process. This issue is addressed in Section \ref{Section::Systematics} where we quantify the systematic errors associated with such a method, and in Section \ref{Section::Simulation} where we apply the same stacking process on simulated lightcurves.

The final stacked lightcurve is from 40 days before the selected flares until 600 days after them, which theoretically allows us to probe apparent speeds as low as 0.5 c for the main flare in the radio core, and 0.8 c for the main flare in the upstream radio knot (at those low apparent speeds, a perturbation would take $\sim 600$ days to reach the next downstream knot).
In order to have a clear picture of a possible variability pattern, each flare is normalized to the strongest one. 
We apply a normalization factor only on fluxes above the full lightcurve median to not alter the flux baseline of the source. The normalized flux for the stacked lightcurve $i$ applied to a data  point $j$ takes this form:
\begin{eqnarray}
\text{If}~ (F_{i,j}-m) &>& 0: \nonumber\\ 
F_{norm, i,j}&=& \frac{F_{max} - m}{F_{max,i} - m} (F_{i,j}-m) + m,
\end{eqnarray}
with $m$  and $F_{max}$ the the median value and the biggest flare respectively in the original lightcurve, $F_{i,j}$ the original flux point in the lightcurve $i$, and $F_{max,i}$ the maximum flux of the lightcurve $i$.
The errorbars are adapted accordingly to keep the same error/flux ratio.

The resulting stacked lightcurve is presented in Figure \ref{Fig::stacked_lightcurve}. At first sight we notice the great dispersion of data points, which is in part due to the duplicate of flares inherent to the stacking method. These flare duplicates have also their fluxes amplified by the normalization process.
But mostly, this dispersion points toward strong stochastic X-ray flux variations of Mrk 421, making it hard to discern a possible intrinsic variability pattern. 

For display purposes, a clearer view is given by rebinning the data. Since the stacked dataset is unevenly sampled with a concentration of points around the stacked flares, we adopt a binning keeping a constant number of data points per bin. The binned data, as well as the RMS dispersion within each bin is shown in Figure \ref{Fig::stacked_lightcurve}. Two excesses at 11 and 23 days after the main flares, and a possible one around 64 days, suggest a post-flare variability pattern. Also the amplitude of these excesses is decreasing with time, which is consistent with adiabatic (expansion) and radiative losses.
In order to evaluate the significance of these suggested features, simulations are required to assess the impact of the various affects such as binning, sampling, and stacking; this is discussed in later sections.

The strong excess seen in the last bin is intriguing. We consider it unlikely associated with the process we want to probe. First, this very long delay seems unlikely associated with the real flow speed of Mrk 421, which is known to show a strongly Doppler boosted radiation. Also the amplitude of such an excess is close to the ones of selected flares, leading to a non-cooling jet over long periods.
It is however possibly highlighting a long-term periodicity of Mrk 421 flares, possibly linked to the accretion disk timescale. 

HBLs are known to be the least powerful blazars and have been associated with a weak accretion mode known as the ``advection dominated accretion flow'' (ADAF).
Approximating the gas flow angular frequency $\Omega$ as the Keplerian angular frequency $\Omega_k$ \citep{Manmoto_1996}, we have
\begin{equation}
\Omega =\left(\frac{G M_\bullet}{r}\right)^{1/2} \frac{1}{r - r_g},
\end{equation}
with the black hole mass $M_\bullet = 1.7 \times 10^{8} M_\odot$ estimated from fundamental-plane-derived velocity dispersion \citep{Woo_2005}, and the associated Schwarzschild radius  $r_g = 5.03 \times 10^{13}$ cm. Then an accretion disk perturbation with an orbital period of 600 days would be located at distance $r = 233~r_g$ from the black hole, which could correspond to the interface between the ADAF and the outer standard thin disk structure \citep{Esin_1997}.


\section{Theoretical models}
\label{Section::model}

\subsection{Multi-Gaussian}

The purpose of the presented model is not to simulate the particle physics processes of a perturbation crossing shocks, such as particle acceleration, cooling, or radiative transfer. Several former studies addressed this approach, via MHD-based and semi-analytic models \citep [e.g.][]{Turler_2000, Mimica_2009, Turler_2011, Fromm_2011, Fromm_2016}. While these models shed light on the shock mechanisms in jets, they would be unfit to statistically probe the existence of a lightcurve pattern induced via multiple shocks due to degeneracies between numerous parameters or inadequately long computation times. Instead, we want to probe a signature of successive shocks with the simplest possible function, and with the maximum physical constraints given by VLBI observations in order to reduce the number of free parameters.

We consider first a general flux baseline $B(t)$ as a linear function, in order to picture a possible long term flux variation of the 640 days' stacked lightcurve not associated with the multiple flares probed:
\begin{equation}
B(t) = f_0 + f_1 t.
\end{equation}

On top of this baseline, a multi-Gaussian function is implemented, a 5-Gaussian function for the radio core flare hypothesis and a 4-Gaussian function for the upstream radio knot hypothesis.
The time-gap $\Delta t_i$ between each peak depends on the free parameter of the apparent speed and the inter-knot gaps measured in the VLBI observation, as expressed in Equation \ref{Eq::TimeGap}. 
The timing of each expected peak $t_i$ can be expressed from the Equation \ref{Eq::TimeGap} by
\begin{equation}
t_i = 7.26 \times 10^2 \frac{(1+z) x_i}{\beta_{app}}  ~\mathrm{days}.
\end{equation}

The spread of the Gaussian $\sigma_{G,i}$ is scaled to the size of the radio knots $R_i$ following this formula:
\begin{equation}
\label{Eq::Sigma}
\sigma_{G,i}  =   \frac{7.26 \times 10^2}{\sqrt{2 \log(2)}} \frac{(1+z) R_i}{\beta_{app}} S + C ~\mathrm{days},
\end{equation}
with $R_i$ the knot radius given in Table \ref{Tab::knots_position} and Figure \ref{Fig::knots_Mrk421} (\textit{Right}).
$S$ and $C$ are scaling factors. The coefficient $\sqrt{2 \log(2)}$ is used to convert the measured size of radio knots expressed in Gaussian FWHM/2 into a standard deviation.
Since we consider a constant apparent speed, the Gaussian spread in days is strictly proportional to a unit of size (see Equation \ref{Eq::width_perturbation}).

Each peak is then defined as
\begin{equation}
P_i(t) = \frac{1}{\sigma_{G,i} \sqrt{2 \pi}} \exp \left[\frac{-(t-t_i)^2}{2 \sigma_{G,i}^2}\right],
\end{equation}

Finally, at constant power, the peaks should have a flux decrease roughly proportional to the volume of the emission zones.
We then express a Gaussian amplitude decrease as
\begin{equation}
A_i = \alpha /\sigma_{G,i}^{3}.
\end{equation}

The full theoretical model, including the baseline is thus given by
\begin{equation}
G_m(t) =  \sum_{i=n}^{5} \left[A_i {P_i}(t)\right] + B(t),
\end{equation}
with $n = 1$ or 2 following the radio core or radio knot hypothesis.

The function $G_m(t)$ contains only 6 free parameters; in order to obtain a realistic model, we constrain the parameter space of some of them.
For a minimal accuracy of the method, we want to be able to probe at least 2 peaks associated with post-flare events in the lightcurve, which sets a minimal apparent speed of $\sim 2$ c considering a secondary peak at the maximum delay of 500 days. The minimal apparent speed is deduced from the closest consecutive knots associated with this delay (we considered a maximum delay shorter than the 600 day probed to be sure to have good resolution of such a peak).
The maximal measured apparent speed in a blazar is $\sim 50$ c, in the jet of  PKS~0805-07 \citep{Lister_2016}.
We consider $\beta_{app} \in [2,70]$.

All the selected X-ray flares in the original lightcurve (considered as the first flare of the sequence) have a duration well below 50 days. We consider this value as a constraint on $\sigma_{G,1}$ from Equation \ref{Eq::Sigma}. In this equation, we assume that the width of the Gaussian cannot grow faster than the width of knots along the jet. Indeed, it is safe to assume that the high-energy shock zone is only a portion of the observed radio knots. Due to the energy loss along the jet propagation, it is likely that this shock zone will not occupy a relatively larger area in the downstream knots. We set the boundaries of $C \in [0,50]$.
So, following the constraint on $\sigma_{G,1}$ and $C$ we set the parameter space of $S \in [0, 3.8]$ for a flare in the radio core and $S \in [0, 1.4]$ for a flare in the upstream radio knot.

The parameter space of all parameters is summarized in Table \ref{Tab::params_model}.

\begin{figure*}[t!]
	\begin{minipage}[b]{0.49\linewidth}
   		\centering \includegraphics[width=8cm]{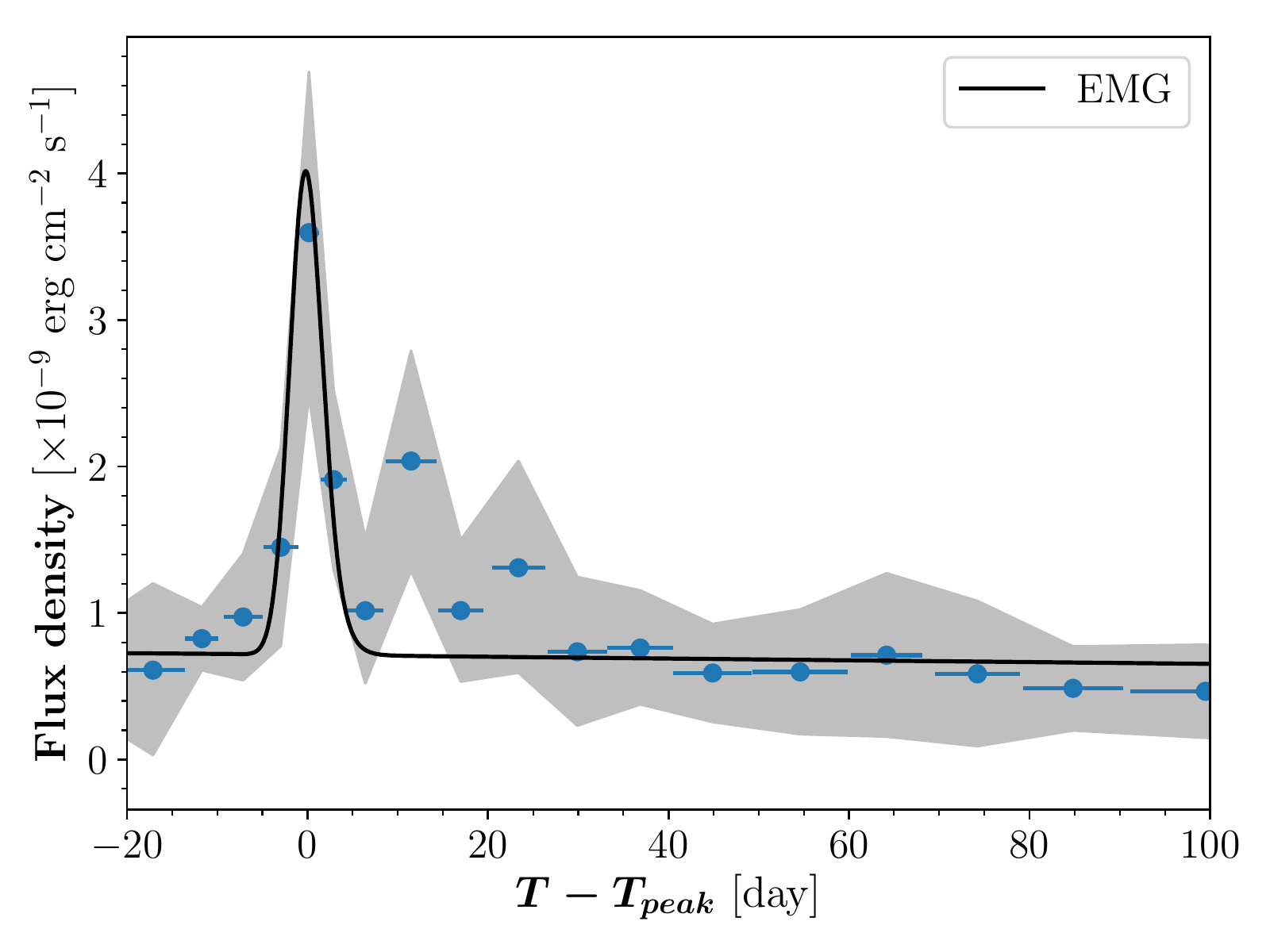}
	\end{minipage}\hfill
		\begin{minipage}[b]{0.49\linewidth}
      \centering \includegraphics[width=8cm]{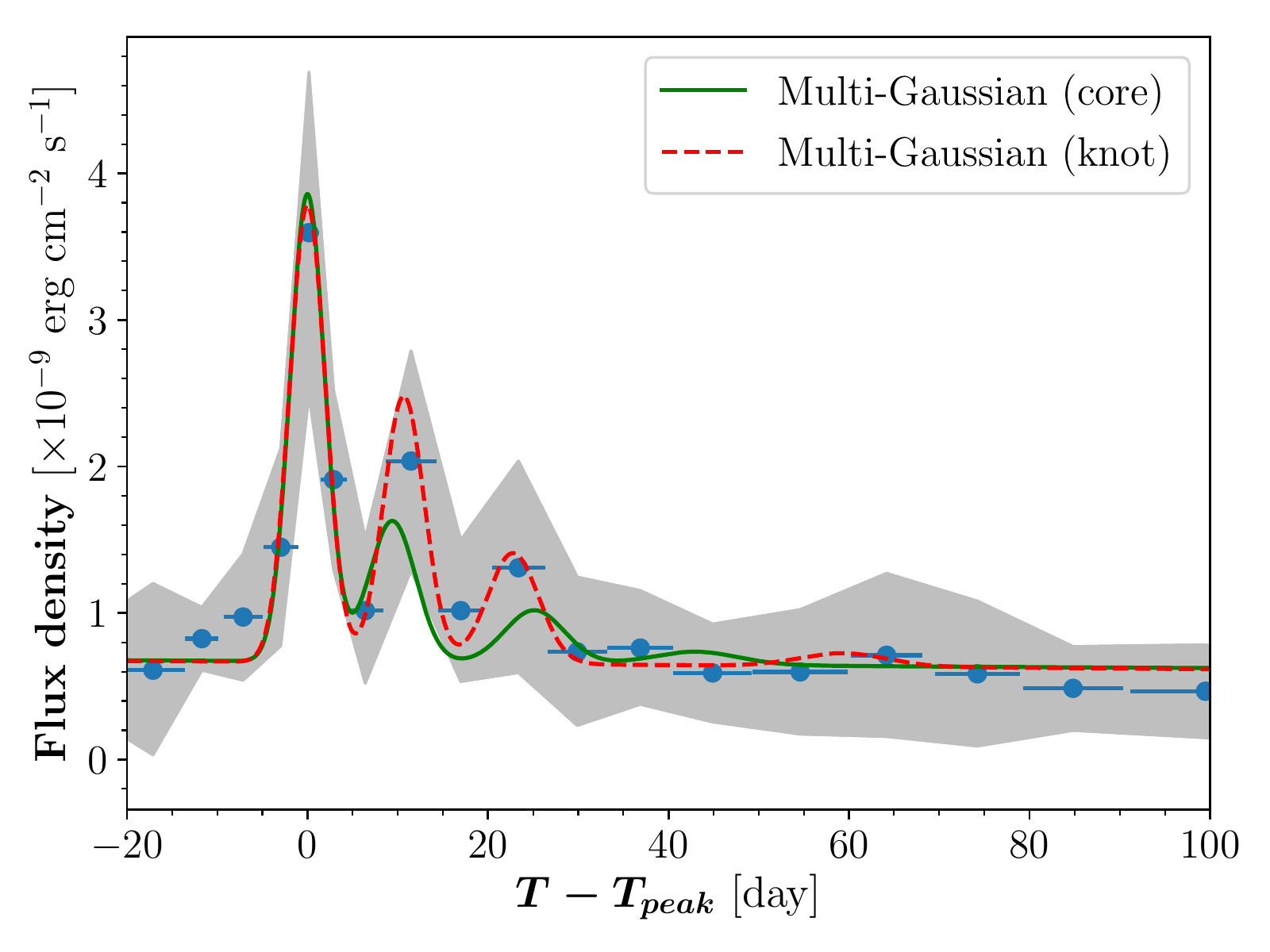}
	\end{minipage}\hfill
	\caption{Comparison of the different flares models; EMG, multi-Gaussian from a main flare in the radio core, and multi-Gaussian from a main flare in the upstream radio knot. The models are represented on top of the stacked lightcurves. For clarity we show a binned dataset, with 18 data points per bin. The grey band is the RMS range associated with the flux dispersion of stacked lightcurves.}
	\label{Fig::fitted-models}
\end{figure*}

\subsection{Exponentially-modified Gaussian}

Blazar flare profiles may present skewness, for which the decay is usually longer than the rise time. Such a skewness is most often modeled by a combination of two exponential functions  \citep[e.g.][]{Abdo_2010, Chatterjee_2012}. We consider a typical flare profile as an exponentially-modified Gaussian (EMG) function, which has similar properties with the two exponential one and the same number of free parameters. The EMG has the specificity to raise as a Gaussian function and decay as an exponential one.
\\
\\

We use the following EMG function expression:
\begin{eqnarray}
EMG(t) = \frac{h \sigma}{\tau} \sqrt{\frac{\pi}{2}} \exp \left(\frac{\sigma^2}{2 \tau^2} - \frac{t - \mu}{\tau} \right) \nonumber \\
\times~\mathrm{erfc}\left[\frac{1}{\sqrt{2}} \left(\frac{\sigma}{\tau} - \frac{t - \mu}{\sigma}\right)\right] + B(t),
\end{eqnarray}
with $h$ the amplitude, $\sigma$ the Gaussian standard deviation, the mean $m = \mu + \tau$ set at 0, and $\tau$ the exponential relaxation time.

As for the multi-Gaussian model, the EMG model takes into account a linear baseline $B(t)$. Hence, the full EMG model has 5 free parameters.

\begin{table}[t]
\centering
\caption{Parameter boundaries applied to the lightcurve models.} 
\label{Tab::params_model}
\begin{tabular}{C|c|c|c}
\tablewidth{0pt}
\hline
\hline
& parameter & boundaries & unit\\
\hline
\decimals
Baseline &$f_0$ & $[0, \infty]$ & erg cm$^{-2}$ s$^{-1}$\\
&$f_1$ & $[-\infty, \infty]$ & erg cm$^{-2}$ s$^{-1}$ day$^{-1}$\\
\hline
&$\alpha$ 	& $[0 , \infty]$ & day$^{4}$ erg cm$^{-2}$ s$^{-1}$\\
Multi- 		&$S_{core}$ & $[0, 3.8]$ & --\\
Gaussian 	&$S_{knot}$ & $[0, 1.4]$ & --\\
&$C$ 		& $[0 , 50]$ & day\\
&$\beta_{app}$ & $[2, 70]$ & c\\
\hline
& $h$ & $[0 , \infty]$ & erg cm$^{-2}$ s$^{-1}$\\\
EMG& $\sigma$ & $[0.5, 50]$ & day \\
& $\tau$ & $[0.5, 50]$ & day \\
\hline
\end{tabular}
\end{table}

\subsection{Model comparison}


The data point dispersion in the stacked lightcurves (associated to intrinsic stochatic variations) have larger amplitudes that the measurement errors associated with each observation. This large data dispersion leads to extremely high values of $\chi^2$, whatever the model used. The models presented do not aim to describe each variation of fluxes in the lightcurve, but look for an intrinsic regular pattern within the stochastic noise. While the fit quality cannot validate a given model by itself, it can however be used to compare the performance of each model.

All the fitted models show excesses above the baseline within a period of 100 days after the stacked flares. Considering only the range where at least one model is above $1\%$ of the baseline, $[t_{flare}-7, t_{flare}+70]$, the fit qualities improve, as well as the relative difference between models (see Table \ref{table::chi2_dof}).

\begin{table}[t]
\centering
\caption{Fitting results for the different proposed models. The first column is for the time range presented in Figure \ref{Fig::fitted-models} while the second column considers the range where at least one model is above $1\%$ of the baseline.} 
\begin{tabular}{c|c|c}
\tablewidth{0pt}
\hline
\hline
 & $\chi^2 /dof$     & $\chi^2 /dof$\\
 & $t\in[-20, +100]$ & $t\in[-7, +70]$\\
\hline
$EMG$		& $7.86 \times 10^5/325$ & $6.05 \times 10^5/233$\\
$G_m,core$	& $7.16 \times 10^5/324$ & $5.26 \times 10^5/232$\\
$G_m,knot$ 	& $6.74 \times 10^5/324$ & $4.83 \times 10^5/232$\\
\hline
\end{tabular}
\label{table::chi2_dof}
\end{table}

The EMG function has the worst $\chi^2$. Although having a visually good representation of the main flare, it does not describe the excesses above the baseline after the flare, contrary to the multi-Gaussian.
Both multi-Gaussian functions, core and knot, are pointing toward a second and third peaks located at $\sim 9 - 11$ and $\sim 22 - 26$ days respectively after the main
flare. 
However the knot scenario is favored with the lowest $\chi^2$ and each of its expected peaks matches the observed flux excesses well.
Thus, in the following we focus on the theoretical model of a main flare from the upstream radio knot.

\subsection{Statistical and systematic uncertainties}
\label{Section::Systematics}

The statistical uncertainties on the fitted model parameters are estimated from the covariance matrix calculation done with the python \texttt{scipy.optimize.curve\_fit} method.\footnote{\url{https://docs.scipy.org/doc/scipy/reference/generated/scipy.optimize.curve\_fit.html}} The data dispersion being much larger than the error associated to each point, the original covariance matrix is scaled to the reduced $\chi^2$ of the best fit to avoid an obvious underestimation of the statistic uncertainties. This process scales the original error bars to match the sample variance of the residuals after the fit.

While being a reasonable method, we raise a warning that the statistical uncertainties estimated this way are likely close to, but not exactly the true ones (e.g. by assuming a normal distribution of the fit residuals).

The way flares are selected in the X-ray lightcurve plays a role in the fitting results, leading to associated systematic uncertainties. We determine the systematic uncertainties of the models parameters by applying different cuts in the flare selection.
As defined in Section \ref{Section:flare_selection} , three cuts are applied to select flares: the flux threshold $FT$, the minimum time gap between two selected flares $\Delta _{Flares}$, and the time range around a given flare where we want a minimum amount of data taken $\Delta t_{data}$.
In order to estimate systematic uncertainties, we consider the effects of applying a much looser and much harder set of cuts. 
The loose cuts select many more flares (13), while the hard ones select fewer (5) but better defined flares.
The different cuts are summarized in Table \ref{table::cuts}.

\begin{table}[t]
\centering
\caption{Cut sets applied to select flares in the X-ray lightcurve. Systematic uncertainties are estimated from the loose and hard cuts.} 
\begin{tabular}{c|c|c|c|c}
\tablewidth{0pt}
\hline
\hline
Cuts & default & loose & hard & unit\\
\hline
$FT$				& $90\%$& $80\%$& $95\%$& flux percentile\\
$\Delta_{Flares}$	& 100	& 75	& 150	& days\\
$\Delta t_{data}$ 	& 10	& 20	& 5		& days\\
\hline
\end{tabular}
\label{table::cuts}
\end{table}

\begin{table}[t]
\centering
\caption{Parameter values for the different studied models with associated systematic uncertainties.} 
\label{tab:params_sys}
\begin{tabular}{c|l|l}
\tablewidth{0pt}
\hline
\hline
parameter & value & uncertainty\\
\hline
\multicolumn{3}{l}{EMG}\\
\hline
$f_0$	 & $(7.13 \pm 0.27 ) \times 10^{-1}$	& $^{+0.40}_{-0.32} \times 10^{-1}$\\
$f_1$ 	 & $(-6.06 \pm 0.90) \times 10^{-4}$	& $^{+2.5}_{-1.7} \times 10^{-4}$\\
$\tau$   & $1.03\pm 0.42$						& $^{+0}_{-0.45}$\\
$\sigma$ & $1.63 \pm 0.21$						& $^{+0}_{-0.25}$\\
$h$      & $3.77 \pm 0.55$						& $^{+0.64}_{-0.71}$\\
\hline
\multicolumn{3}{l}{core-flare model} \\
\hline
$f_0$			& $(6.67 \pm 0.28 ) \times 10^{-1}$	& $^{+0.36}_{-0.25} \times 10^{-1}$\\
$f_1$ 			& $(-4.80 \pm 0.91) \times 10^{-4}$	& $^{+2.1}_{-1.4} \times 10^{-4}$\\
$\alpha$ 		& $(-1.09 \pm 0.25) \times 10^{2}$		& $^{+0.48}_{-0} \times 10^{2}$\\
$S$ 			& $(4.1 \pm 0.9) \times 10^{-1}$		& $^{+1.3}_{-1.3} \times 10^{-1}$\\
$C$ 			& $1.56 \pm 0.13$						& $^{+0}_{-0.25}$\\
$\beta_{app}$ 	& $30.3 \pm  1.6$						& $^{+3.9}_{-1.2}$\\
\hline
\multicolumn{3}{l}{knot-flare model} \\
\hline
$f_0$			& $(6.63 \pm 0.27 ) \times 10^{-1}$	& $^{+0.36}_{-0.23} \times 10^{-1}$\\
$f_1$ 			& $(-4.69 \pm 0.90) \times 10^{-4}$	& $^{+1.9}_{-0.7} \times 10^{-4}$\\
$\alpha$ 		& $(-1.22 \pm 0.26) \times 10^{2}$		& $^{+0.53}_{-0} \times 10^{2}$\\
$S$ 			& $(2.4 \pm 0.6) \times 10^{-1}$		& $^{+1.1}_{-0.5} \times 10^{-1}$\\
$C$ 			& $1.57 \pm 0.15$						& $^{+0.03}_{-0.39}$\\
$\beta_{app}$ 	& $44.6 \pm  1.2$						& $^{+3.8}_{-0.3}$\\
\hline
\end{tabular}
\end{table}

The systematic loose cuts uncertainties for each parameter are calculated as $\Delta_{sys, loose} = \mathrm{loose} - \mathrm{default}$.
The same is applied for hard cuts. If loose and hard cuts values are not bracketing a default parameter value, only the larger $\Delta_{sys}$ is taken into account.
The default parameter values and the systematic uncertainties for the different models are given in Table \ref{tab:params_sys}.

These two alternative cut sets do not impact the favored interpretation of the strongest flares originating from the upstream radio knot. Indeed, the knot-flare model always has the lowest $\chi^2$ value, whatever the cut choice.

\section{lightcurve simulation}
\label{Section::Simulation}

The significance of the knot-flare scenario against the null hypothesis can be estimated via comparisons with multiple realistic simulated lightcurves of Mrk 421. 
By applying the exact same method on simulated lightcurves, one can estimate the probability that the observed post-flare variability pattern is from pure stochastic noise.

The conditions we want to fulfill for the simulated lightcurves compared to the original one are:
\begin{itemize}
\item[-] Similar power spectrum density (PSD)
\item[-] Similar time sampling
\item[-] Similar flux distribution
\end{itemize}

\subsection{Power spectrum density}

The Swift-XRT PSD is produced using the LombScargle package of Astropy.\footnote{\url{http://docs.astropy.org/en/stable/stats/lombscargle.html}} 
The frequency range considered to build the PSD is delimited by the total lightcurve length, $\nu_{min} = 1/T$ with $T$ the 13.3-year span of the total lightcurve, and the Nyquist frequency defined as $\nu_{max} =N/(2 T)$ with $N$ the number of data points \citep[e.g.][]{Uttley_2002}.

\begin{figure}[t!]
\begin{center}
	\includegraphics[width= 0.5\textwidth]{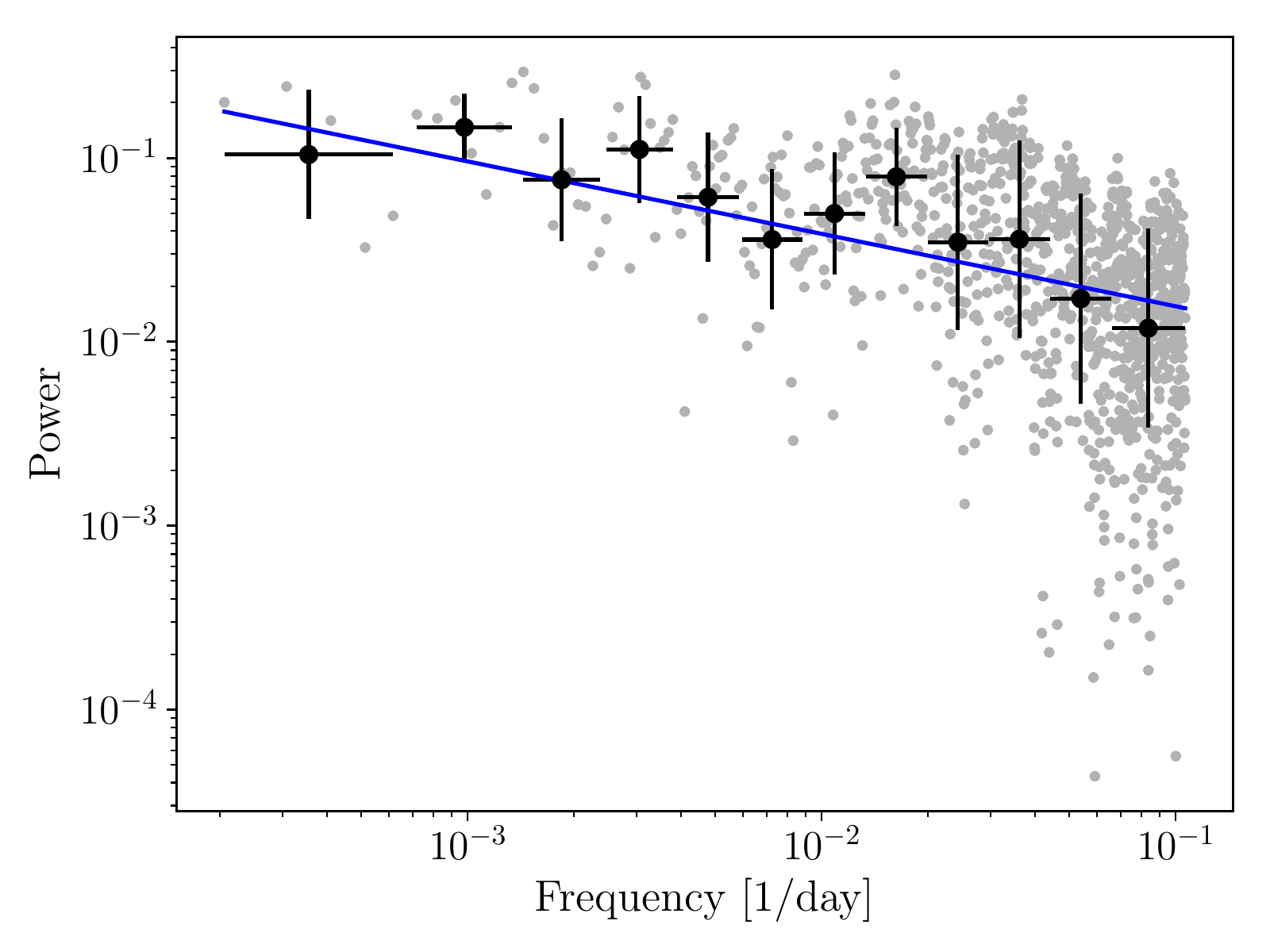}
 	\caption{\textit{Swift}-XRT PSD fitted by a power-law function. Black points are logarithmically binned data.}
 		\label{Fig::Power_spectral_density_binned}
 \end{center}
\end{figure}

The PSD index is  extracted from a power-law fit, with a best value of $\eta = 1.35 \pm 0.01$ ($P_\nu \propto \nu^{-\eta}$).
The power-law function has a good fit with $\chi^2_{red} = 0.39$ for the logarithmically binned PSD shown in Figure \ref{Fig::Power_spectral_density_binned}.

\subsection{Sampling}
\label{Section::Sampling}

A simulated lightcurve is produced considering power-law noise with the index $\eta$ by the \texttt{astroML.time\_series} \texttt{.generate\_power\_law} tool \footnote{\url{http://www.astroml.org/modules/generated/astroML.time_series.generate_power_law.html}} \citep{astroML}, based on the method developed by \cite{Timmer_1995}.

In order to avoid the red noise leak (transfer of variability power from the low to high frequencies due to the finite length of observations), we simulate lightcurves 100 times larger than the observed one, then clip it to the original length.
Also, the \textit{Swift}-XRT dataset is far from evenly sampled, mostly due to the observations being taken as "Targets of Opportunity".
Since having a different sampling in simulated lightcurves would bias a fair statistical test, we re-sample the simulated lightcurves by taking the interpolated fluxes corresponding to each observing date of \textit{Swift}-XRT.

\subsection{Producing a realistic lognormal distribution}

Mrk 421 is known to show a lognormal flux distribution from radio to very-high energies \citep{Tluczykont_2010, Sinha_2016,Kushwaha_2017}.

\begin{figure}[t]
\begin{center}
	\includegraphics[width= 0.5\textwidth]{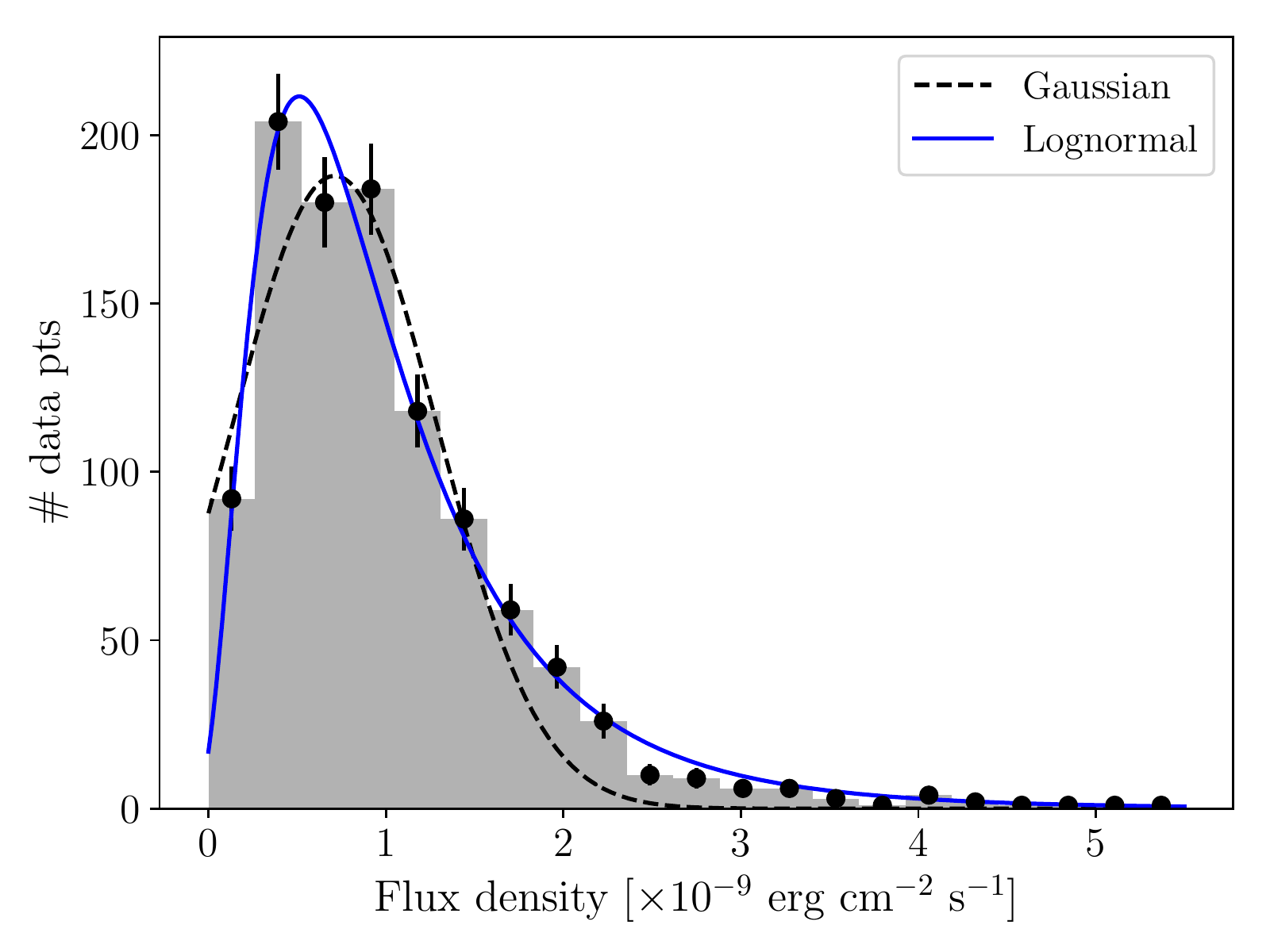}
 	\caption{Comparison between Gaussian and lognormal functions fitted to the \textit{Swift}-XRT flux distribution.}
 		\label{Fig::Flux_distribution}
 \end{center}
\end{figure}

We confirm this behavior in our \textit{Swift}-XRT dataset by testing a lognormal against a normal distribution hypothesis. Both have 19 degrees of freedom. The reduced $\chi^2$ of the lognormal function shows a better fit, with $\chi^2_{red,Lognorm} = 1.58$ and $\chi^2_{red,Gauss} = 4.85$.
Assuming a usual p-value acceptance limit of $0.05$, the lognormal function is accepted with $P_{Lognorm} = 5.2 \times 10^{-2}$, while the Gaussian assumption is strongly rejected with $P_{Gauss} = 1.3 \times 10^{-11}$.
These two fits are shown in Figure \ref{Fig::Flux_distribution}.

\begin{figure}[t]
\begin{center}
	\includegraphics[width= 0.5\textwidth]{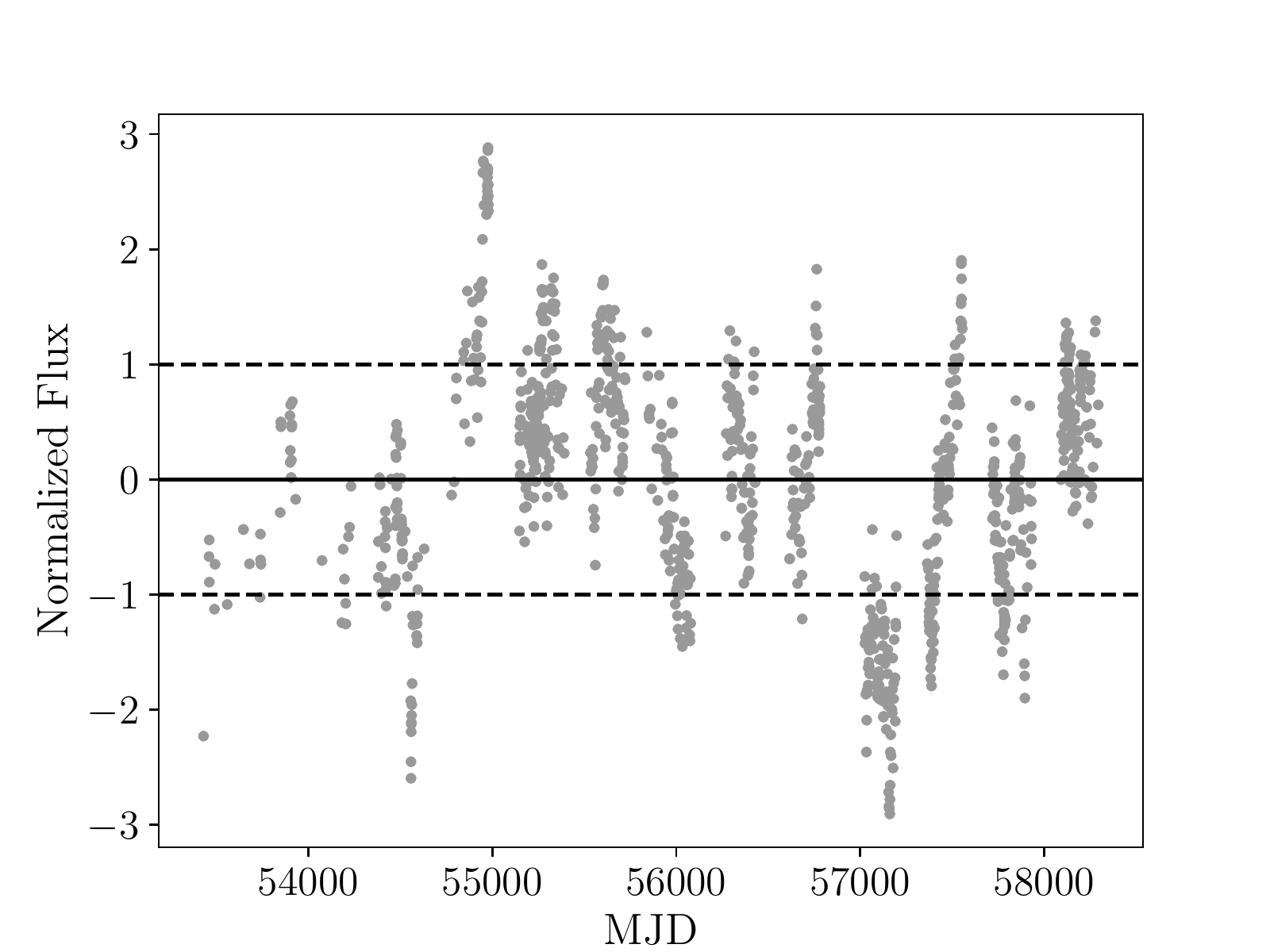}
 	\caption{Example of a re-sampled and normalized simulated lightcurve. The black solid and dashed lines are the mean and the variance of the dataset respectively.}
 		\label{Fig::Simulated_lightcurve_Norm}
 \end{center}
\end{figure}

Before adjusting the simulated lightcurves to the one with a realistic distribution, we need to normalize their variance $V(\Phi(t))$ to 1 and mean value $\langle\Phi(t)\rangle$ to 0. An example of such a re-sampled and normalized lightcurve $\Phi_{sim,norm} (t)$ is given in Figure \ref{Fig::Simulated_lightcurve_Norm}.

Then the distribution can be transformed to lognormal following the equation
\begin{equation}
\Phi_{sim,LN}(t) = \exp\left((\Phi_{sim,norm} (t) \times a) + b\right),
\end{equation}
with $a = \sigma_{sim}$ and $b = \mu_{sim}$ of the normally distributed logarithm $\log(\Phi_{sim,LN}(t))$. This comes from the fact the mean and variance of $\Phi_{sim,norm} (t)$ are 0 and 1 respectively.

These two parameters a and b can be observationally constrained considering that observed and simulated lightcurves should have similar mean value as well as similar variability amplitude $F_{var}$.

The variability amplitude, as defined by \cite{Rodriguez_1997}, is expressed as
\begin{equation}
F_{var} = \frac{\sqrt{V\left(\Phi(t)\right) - \Delta^2 }}{\langle\Phi(t)\rangle},
\end{equation}
with $\Delta^2$ the mean square value of uncertainties. 
At this point the simulated dataset does not have yet associated uncertainties, so $F_{var,sim}$ can be expressed only from the variance and the mean. They have for a lognormal distribution these forms:
\begin{equation}
\label{Eq::meanPhi}
\langle\Phi(t)\rangle = \exp(\mu + \sigma^2/2)
\end{equation}
\begin{equation}
V\left(\Phi(t)\right) = (e^{\sigma^2} - 1) \langle\Phi(t)\rangle^2
\end{equation}
So $\displaystyle{F_{var,sim}^2 = e^{\sigma_{sim}^2} - 1} $.

Knowing that the coefficient $a = \sigma_{sim}$ and we want similar observed and simulated $F_{var}$, we can write $a$ as
\begin{equation}
a = \sqrt{\left(\log(F_{var,obs}^2) + 1 \right)}.
\end{equation}

Then following Eq. \ref{Eq::meanPhi}, and given the assumption of similar observed and simulated mean values $\langle\Phi(t)\rangle$, the coefficient $b$ takes the form 
\begin{equation}
b = \langle\Phi_{obs}(t)\rangle - a^2/2.
\end{equation}

From the amplitude variability $F_{var, obs} = 0.68$, we deduce the values $a = 0.62$ and $b = 3.69$.

Instead of having similar $F_{var}$, one can choose to have similar median values between observed and simulated $\Phi(t)$. The median value of a lognormal distribution is defined as
\begin{equation}
median(\Phi(t)) = e^\mu.
\end{equation}

Then, we have the corresponding values of $a = 0.52$ and $b = 3.74$. 

Finally, by directly doing a Gaussian fit to $\log(\Phi_{obs}(t))$, we obtain the coefficients $a =0.61\pm0.03$ and $b=3.79\pm0.03$.

We can explain the differences between these three estimations by considering that the \textit{Swift}-XRT lightcurve does not exactly follow a lognormal distribution, and $F_{var, obs}$ has intrinsic uncertainties \citep{Vaughan_2003}.

For the simulated lightcurves we consider the middle ground between these three estimations by taking the average values of $a = 0.59$ and $b = 3.74$.

\begin{figure}
\begin{center}
	\includegraphics[width= 0.5\textwidth]{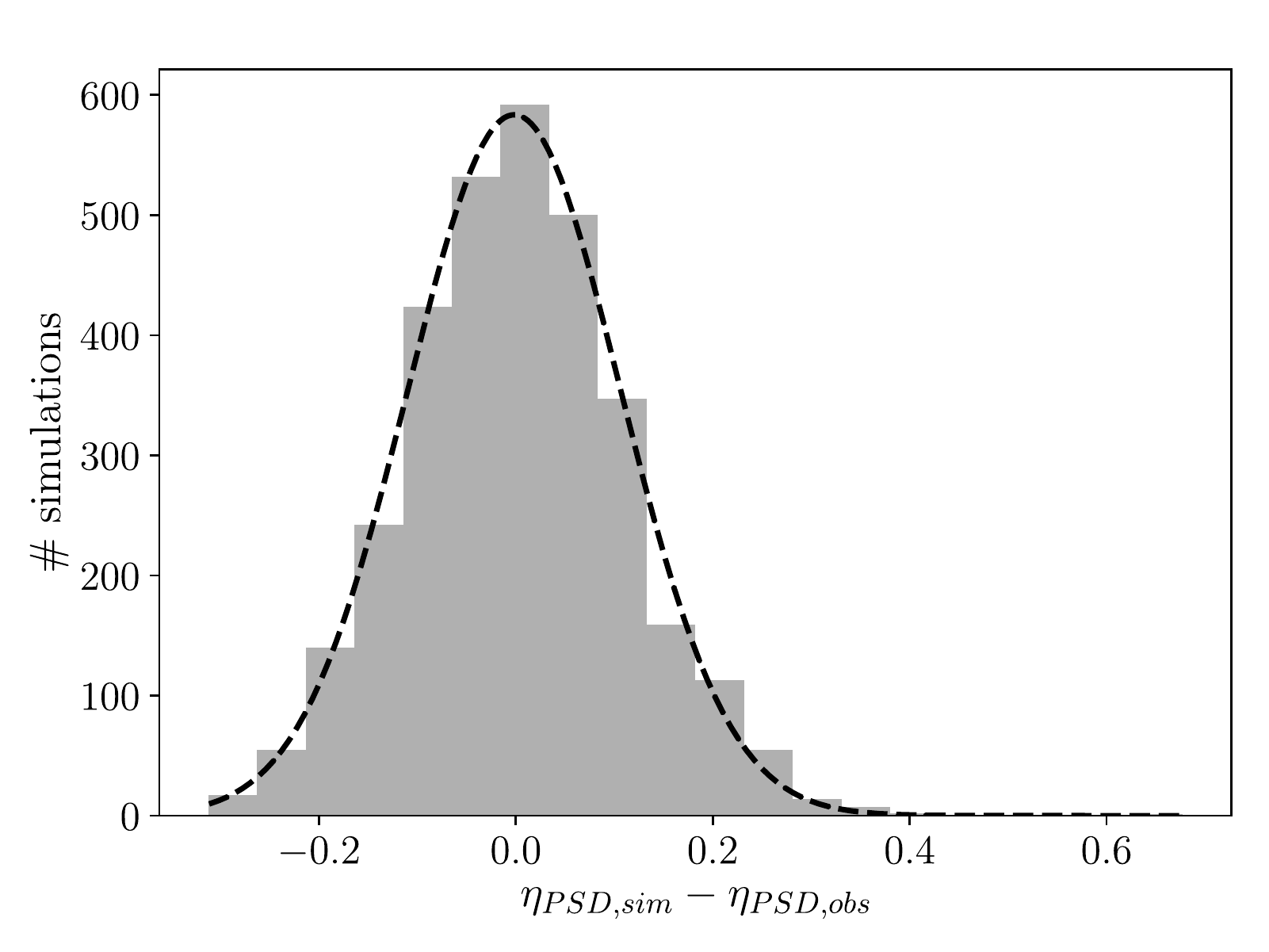}
 	\caption{Distribution of the differences between reconstructed and original PSD indexes for 3200 simulated lightcurves, after applying the correction factor of $\Delta \eta_{PSD} = +3.3 \times 10^{-2}$. The distribution has a mean value of $(-1.7 \pm 1.8) \times 10^{-3}$ with a standard deviation of $(1.08 \pm 0.01) \times 10^{-1}$.}
 		\label{Fig::PSD_Index}
 \end{center}
\end{figure}

\subsection{Simulated errors}

The simulated errors on fluxes should also be realistic. 
We notice the absence of significant correlation between the \textit{Swift}-XRT fluxes and associated uncertainties, with a Pearson correlation coefficient of $r = 0.0059$ and the p-value $P= 0.85$.
Since the simulated lightcurves have the same number of data points as the original one, we simply associate each of the simulated lightcurves with the observed uncertainties randomly shuffled. This method ensures the exact same distribution of uncertainties for all simulations.
Finally, each point is randomly projected following a normal distribution, with its standard deviation given by the error bar.

\subsection{Checking the simulated lightcurves}

After all the processes described above, the simulated lightcurves have PSD indexes which differ from the original one. The distribution of the reconstructed PSD index of a large number of simulations ($\eta_{PSD,sim} - \eta_{PSD,obs}$) is checked by fitting this distribution with a Gaussian. The resulting mean value of $(-9.6 \pm 1.5) \times 10^{-3}$  highlights a significant bias, about 6 sigma, that simulated lightcurves show on average lower PSD indexes.

We correct this bias by iteratively testing various values of $\eta_{PSD}$ used to reconstruct the lightcurves, and stop the iteration when converging towards a $< 1$ sigma discrepancy, corresponding to a correcting factor of $\Delta \eta_{PSD} = +3.3 \times 10^{-2}$ giving consistent results between observed and simulated indexes, as shown in Figure \ref{Fig::PSD_Index}.

\begin{figure}[t!]
\begin{center}
	\includegraphics[width= 0.5\textwidth]{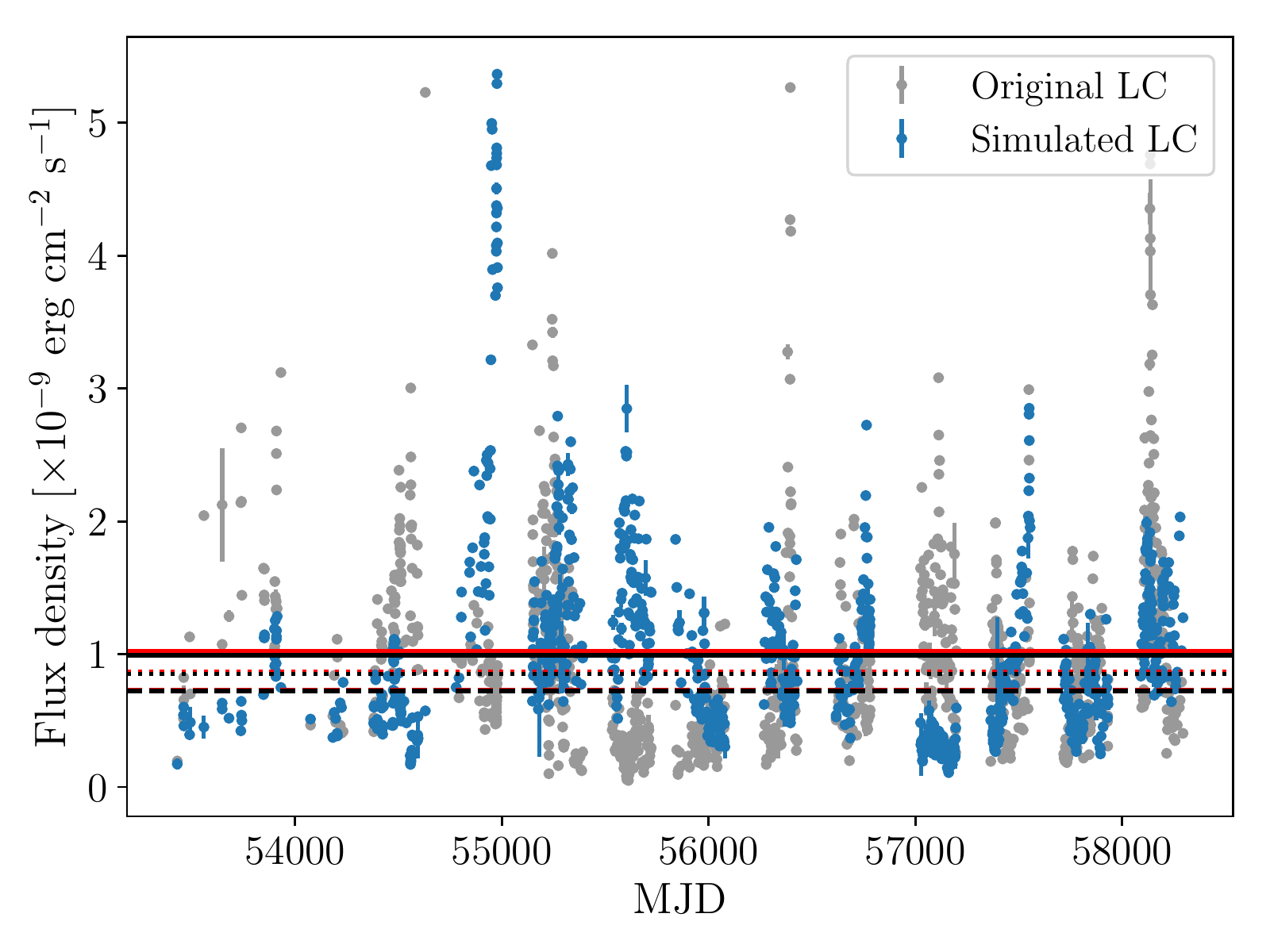}
 	\caption{Example of a simulated lightcurve passing all the checks, compared to the original one. The solid, dashed, and dotted lines are the mean, median, and standard deviation respectively. Black lines are for the original lightcurve while red lines are for the simulated one.}
 		\label{Fig::Simulated_vs_obs_lightcurves}
 \end{center}
\end{figure}

The reconstructed lightcurves being based on Monte-Carlo simulations, with potential strong alterations due to the re-sampling process, we perform further checks to ensure that all simulations are realistic enough to be used for our statistical comparison.
Simulated lightcurves are considered good when they have a reconstructed PSD index and a lognormal distribution ($\mu$ and $\sigma$) within 3 standard deviation of the ones of the original.
An example of such a simulated lightcurve passing all the checks is shown in Figure \ref{Fig::Simulated_vs_obs_lightcurves}.

\begin{figure*}
	\begin{minipage}[t]{0.3\linewidth}
   		\centering \includegraphics[width=6cm]{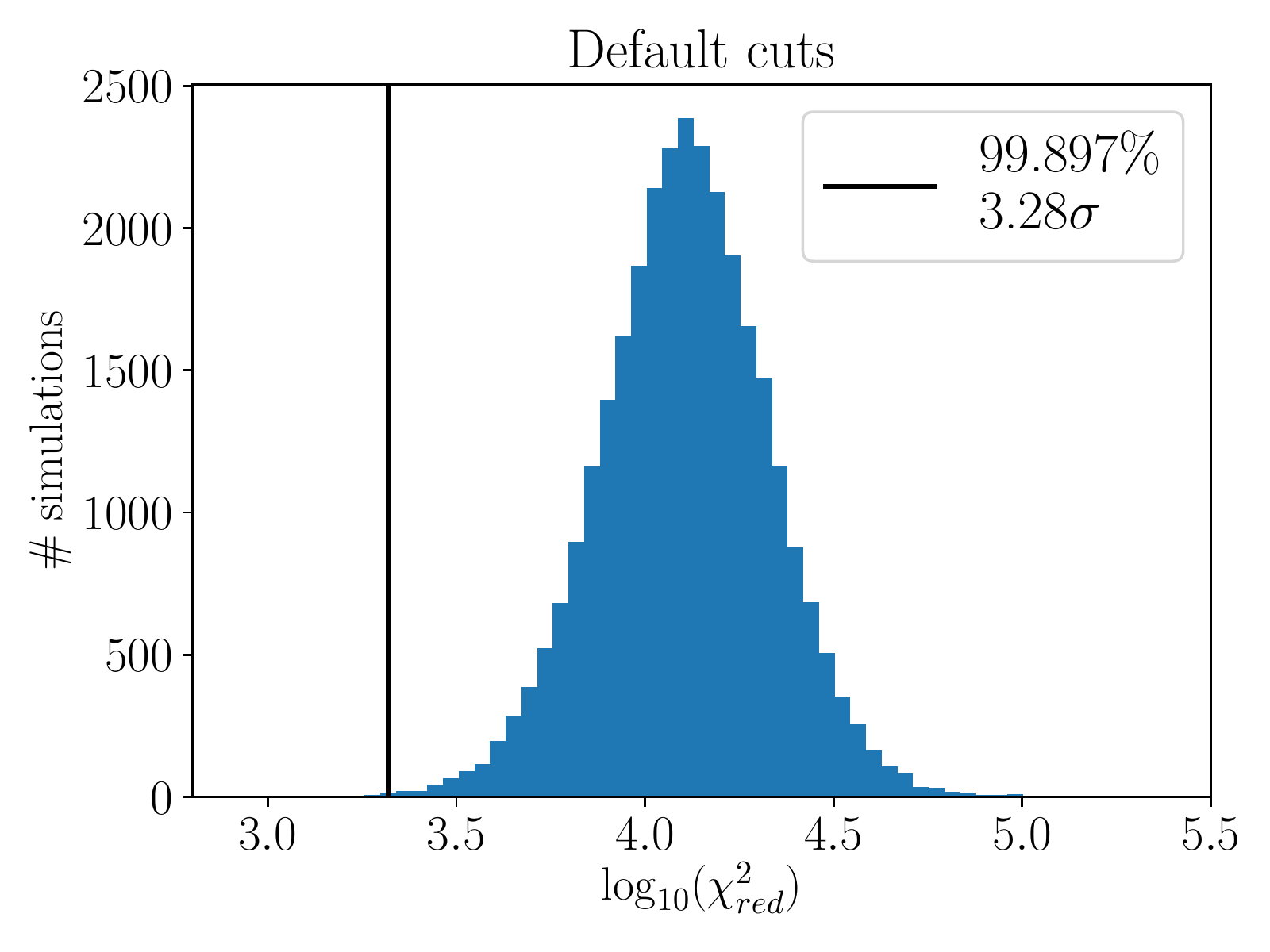}
	\end{minipage}\hfill
		\begin{minipage}[b]{0.3\linewidth}
      \centering \includegraphics[width=6cm]{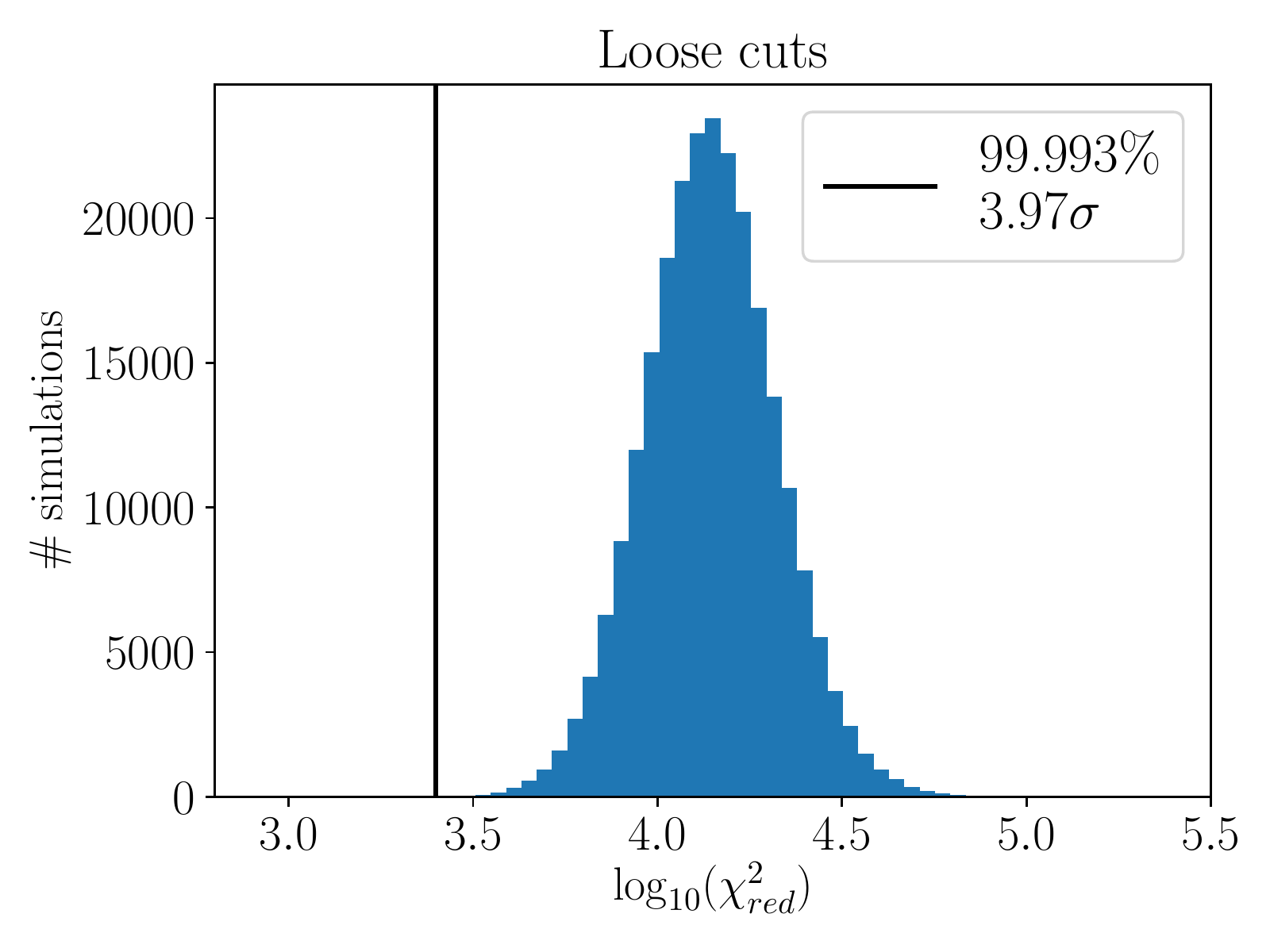}
	\end{minipage}\hfill
	\begin{minipage}[b]{0.3\linewidth}
      \centering \includegraphics[width=6cm]{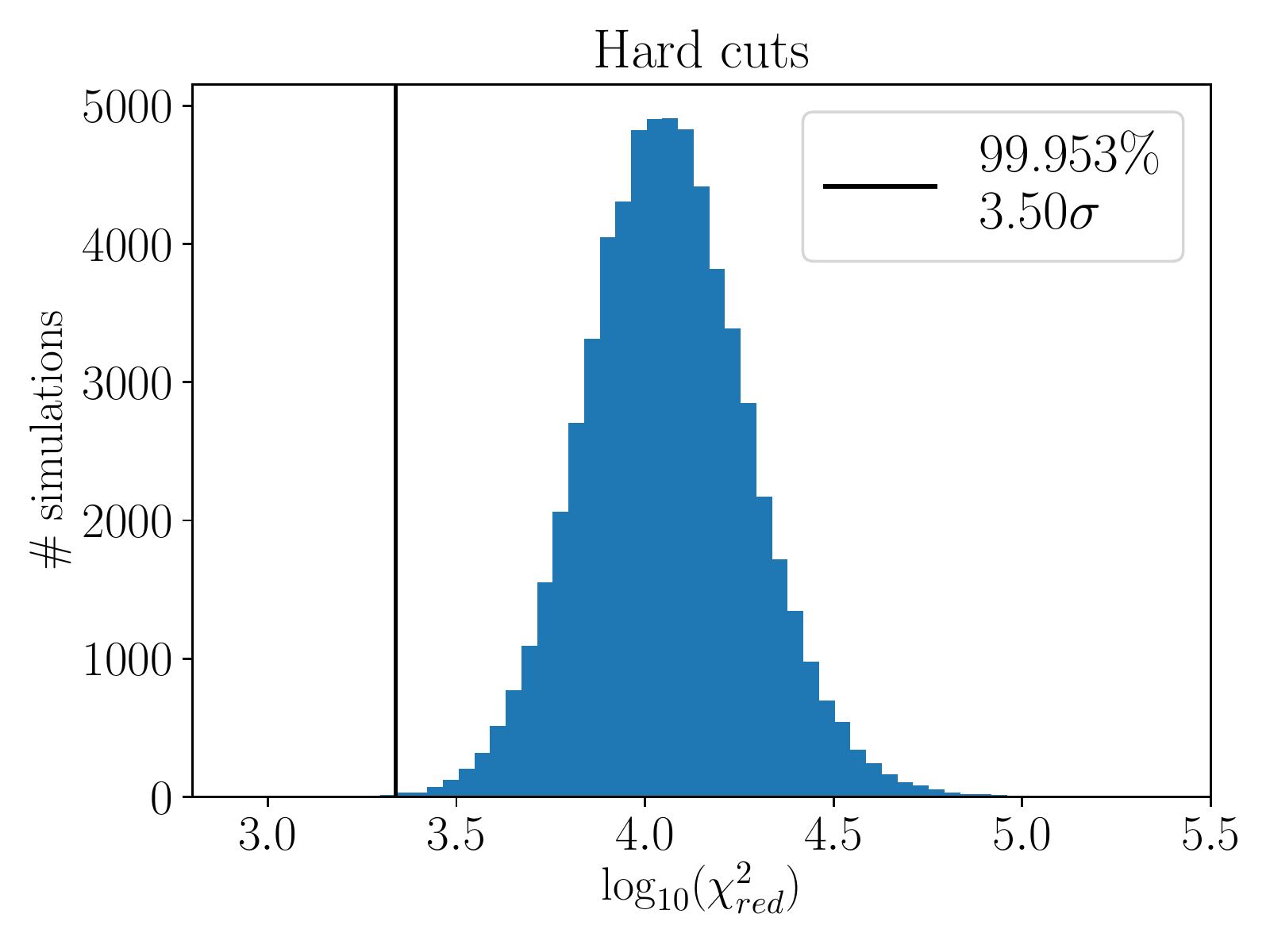}
	\end{minipage}\hfill
	\caption{Distribution of the multi-Gaussian fit results ($\chi^2_{red}$) on large samples of simulated lightcurves, considering the Default, Loose, and Hard flare selection cuts. The black lines are the result for the original \textit{Swift}-XRT lightcurve. The intrinsic multi-Gaussian post-flare pattern of Mrk~421 is validated above a $3 \sigma$ level against stochastic fluctuations for all 3 sets of cuts.}
	\label{Fig::Results}
\end{figure*}

\subsection{Bias of "Target of Opportunity" observations}

As discussed in Section \ref{Section::Sampling}, the fact that \textit{Swift}-XRT mostly observes Mrk~421 as a Target of Opportunity (ToO) introduces a non-even sampling of the dataset, which is fully considered in the simulated lightcurves by the re-sampling process.
However it induces another bias which cannot be easily simulated. Working in response to a ToO means better sampled observations when a flare is occurring. Following the ToO criteria, denser observations are taken when a flux reaches a given threshold defined by the observers.

This is not the case for simulated lightcurves, which leads to fewer and weaker flares passing the selection cuts on average.
It has the effect to reduce the data dispersion of fit residuals in stacked simulated lightcurves, and so leads to lower reduced $\chi^2$, which biases the statistic test in favor of the simulations.

This bias can be taken into account by applying a selection cut on the simulated lightcurves based on the minimum number and minimum flux average $M_{F,flares}$ of selected flares. 
By working on a large number of simulations, we adjust these two cuts to produce results as close as possible to the ones of the original lightcurve.
We do not want the simulations to have higher number of selected flares and $M_{F,flares}$ on average, which would  bias the statistic test in the other way. Keeping these average values slightly below the original ones ensures having a conservative estimate of the probed model significance. These cut values are shown in Table \ref{Tab::final_cuts}.

\begin{table}
\renewcommand{\thetable}{\arabic{table}}
\centering
\caption{Post-simulation cuts to select lightcurves with enough selected flares and sufficiently high average fluxes $M_{F,flares}$ to ensure a fair comparison with the \textit{Swift}-XRT dataset. These cuts are adjusted for the three flares selections described in Section \ref{Section::Systematics}.} 
\label{Tab::final_cuts}
\begin{tabular}{l|l|l|l}
\tablewidth{0pt}
\hline
\hline
						& Default  & Loose  		& Hard \\
\hline
Original dataset &&&\\
\hline
nb flares 				& 6    		& 13      	 	& 5 \\
$M_{F,flares}$* 		& 3.68    	& 2.74       	& 4.02 \\
\hline
Simulations &&&\\
\hline
Cut (nb flares)			& $\geq 4$  & $\geq 11$		& $\geq 4$	 \\
Cut ($M_{F,flares}$)*  	& $\geq 3.3$& $\geq 2.5$	& $\geq 3.5$	 \\
\hline
\multicolumn{4}{l}{\footnotesize{* Fluxes in $10^{-9}$ erg cm$^{-2}$ s$^{-1}$}}
\end{tabular}
\end{table}

\section{Results and discussion}
\label{Section::Discussion}

\subsection{Significance of the multiple-shock scenario}

From the simulations described in the previous section, we can now provide a fair comparison with the original dataset.
At the end, only a small portion of the simulated lightcurves ($\sim 1/10 - 1/20$) are passing all the cuts to be considered realistic enough for a statistical test. Several millions lightcurves are then produced to have enough statistics.  The fraction of simulated lightcurves $f_{sim}$ having a knot-flare model fit worse than the one of the original dataset can be converted to the significance of the intrinsic post-flare pattern result against stochastic fluctuations. We use this expression:

\begin{equation}
\sigma_{result} = \mathrm{erf}^{-1}(f_{sim}) \sqrt{2}.
\end{equation}

Due to the varying number of degrees of freedom in each stack of simulated lightcurves,  the reduced $\chi^2$ is used as an estimator of the fit quality. Also, the post-flare series probed are mostly occurring in a small temporal region of the 640-day stacked lightcurves. Comparing the $\chi^2_{red}$ on these 640 days would give too much importance to the baseline fit quality rather than the probed post-flare scenario.
Hence we consider the $\chi^2_{red}$ only for the time range between the first and the last Gaussian. This time range is defined between the first and the last data point where the multi-Gaussian model is $1\%$ above the baseline.
The $\chi^2_{red}$ associated with Default, Loose, and Hard cuts are $\chi^2_{red,Default} = 2.09 \times 10^3$, $\chi^2_{red,Loose} = 2.50 \times 10^3$, and $\chi^2_{red,Hard} =  2.18 \times 10^3$ respectively.

The significances of the knot-flare scenario against a stochastic process from lightcurve simulations for the three set of cuts are between 3.28 and 3.97 sigma (see Figure \ref{Fig::Results}). The biggest significance of 3.97 sigma is found for the Loose cuts.  The decrease of fit quality of the \textit{Swift}-XRT data associated with the noise induced by 13 selected flares in the Loose cuts is less than the average one of simulations, leading to a better significance than the Default cuts with 6 selected flares.
This suggests that the intrinsic post-flare pattern is also present in weaker flares.

\subsection{Characterization of the jet and perturbation}

The deduced apparent flow speed of the VLBI jet of Mrk 421 of $\beta_{app} = 44.6^{+4.0}_{-1.2}$ gives a physical constraint on the maximum angle with the line of sight $\theta$ as 
\begin{eqnarray}
\label{Eq::theta}
\theta &<& 2 \arctan(1/\beta_{app}),
\end{eqnarray}
leading to $\theta \leq 2.69$ deg, when considering a $90 \%$ confidence level limit.

The jet Doppler and Lorentz factors can be both expressed in functions of the apparent speed and the angle with the line of sight, following these formulas:

\begin{eqnarray}
\delta &=& \sqrt{1 - \left(\frac{\sin \theta}{\beta_{app}} + \cos \theta \right)^{-2}}  \left(1 + \frac{\beta_{app}}{\tan \theta }\right)\\
\label{Eq::lorentz_bapp}
\Gamma &=&  \frac{1}{\sqrt{1 - \left(\displaystyle{\frac{\sin \theta}{\beta_{app}}} + \cos \theta \right)^{-2}}}
\end{eqnarray}

This parameter space can have an additional constraint from the jet opening angle of Mrk 421. Indeed a canonical relation links the apparent jet full opening angle $\alpha_{app}$ with the Lorentz factor, which can be expressed as 
\begin{equation}
\Gamma = \frac{2 \rho}{\alpha_{app} \sin\theta}.
\label{Eq::Lorentz_opening}
\end{equation}

This equation can be seen as an approximation of relativistic jet gas dynamics,  where the Lorentz factor depends on the opening angle and the ratio of pressure between the jet core and the external medium $P_{ext}/P_0$ \citep{Daly_1988,Jorstad_2005}.
The deduced value of $\rho = 0.17 \pm 0.08$ from multiple jet radio VLBI measurements (opening angle, apparent speed, and variability) by \cite{Jorstad_2005} leads to $P_{ext}/P_0 \simeq 1/3$, which corresponds to a case where jets naturally form standing recollimation shocks \citep{Daly_1988}, fully consistent with the probed multiple-shock scenario.

\begin{figure}
\begin{center}
	\includegraphics[width= 0.5\textwidth]{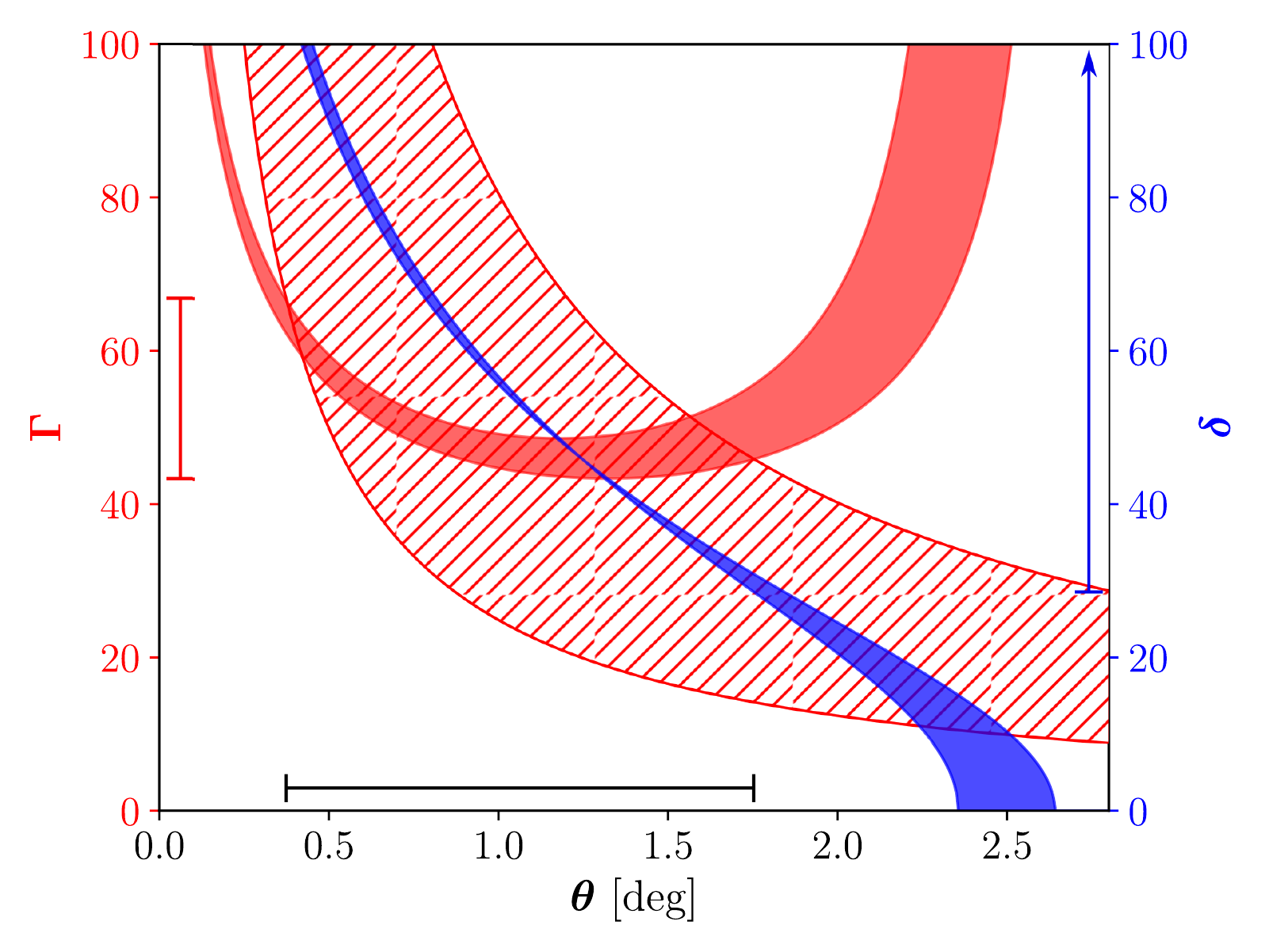}
 	\caption{Lorentz factor (red) and Doppler factor (blue) as a function of the angle with the line of sight $\theta$, the plain bands are calculated given the uncertainty on $\beta_{app}$, while the red-hashed band is calculated given the uncertainties on $\rho$ and $\alpha_{app}$. The segments are showing the likely range for each parameter.}
 		\label{Fig::Doppler_Lorentz}
 \end{center}
\end{figure}

The apparent opening angle can be deduced from the slope $\phi$ of the linear fit shown in Figure \ref{Fig::knots_Mrk421} (\textit{right}) as $\alpha_{app} = 2 \arctan(\phi)$.
Thus, as shown in Figure \ref{Fig::Doppler_Lorentz}, the system can be resolved within the parameter ranges $\theta \in [0.38 - 1.8]$ deg, $\Gamma \in [43 - 66]$, and $\delta \geq 31$. 

This Doppler factor lower limit is relatively high compared to previous estimations of Mrk 421 from SED modeling with $\delta \sim 20-25$ \citep{Katarzynski_2003, Aleksic_2015, Balokovic_2016}, but is consistent with the range of $\delta \in [15-35]$ deduced by \cite{Tavecchio_1998} from broadband SED parametrization. We can note that the maximum Doppler value is quite difficult to estimate from SED models due to the known degeneracy between the parameters.

The width of the multiple Gaussian given by Equation \ref{Eq::Sigma} provides valuable information to constrain general features of the perturbation crossing the shocks. 
In the following, we assume that particle acceleration and cooling times are shorter than the shock crossing time of a perturbation. This assumption implies that the duration of a flare is roughly equal to the duration of the perturbation crossing a shock.

We consider that each Gaussian peak $P_i$ is defined as the convolution product of a Gaussian perturbation $P_p$ crossing a Gaussian shock $P_s$. The standard deviations can then be written as

\begin{equation}
\sigma_{G,i} = \sqrt{\sigma_{p}^2 + \sigma_{s,i}^2}.
\end{equation}

The width of the perturbation, expressed as the Gaussian FWHM takes the form
\begin{equation}
W_p = 2 \sqrt{2 \log{(2)}}~ \sqrt{\sigma_{G,i}^2 - \sigma_{s,i}^2} ~\frac{c \beta_{app}}{1+z}.
\label{Eq::width_perturbation}
\end{equation}

The shock standard deviation $\sigma_{s,i}$ can be constrained; from 0 for a perpendicular shock with no width, to an upper limit at the size of the radio knots:
\begin{equation}
0 \leq \sigma_{s,i} \leq  \frac{R_{i}}{\sqrt{2 \log{(2)}}} ~ \frac{1+z}{c\beta_{app}}.
\end{equation}

Then we can determine the perturbation width from the boundaries on $\sigma_{s,i}$.  The first shock gives the strongest constraints, leading to a value of $W_p = 3.9^{+1.5}_{-3.0} \times 10^{17}$ cm, taking into account the uncertainties on the knot measured radius and fit parameters (statistical and systematic).
The co-moving intrinsic width can be written as
\begin{equation}
\label{Eq::intrinsic_width}
W_{p,int} = \frac{W_p}{\Gamma \sin\theta}.
\end{equation}

Given the values of $\Gamma$ and $\theta$ deduced above, the perturbation intrinsic width lies within the range $W_{p,int} \in [0.43 - 19] \times 10^{17}$ cm.

\subsection{A new look on Mrk 421 emission scenarios}

Mrk 421 is known to present a flux-flux correlation between X-rays and gamma rays, specifically strong in the VHE regime ($E > 100$ Gev). This correlation has been observed in flares and short timescale variability \citep{Fossati_2008,Horan_2009,Acciari_2011}  as well as in period of months to years \citep{Acciari_2014}. It was also noticed that this correlation extends even to the lowest observed fluxes of Mrk 421 \citep{Balokovic_2016}.
It indicates that gamma rays and X-rays are coming from the same emission zones, whatever the activity state of the source. In the context of the present study, it means that the flaring gamma-ray emission zones are located inside the radio knots.

Mrk 421 is also known to present strong and fast outbursts in X-rays and gamma rays, on timescales of $ \sim 15$ minutes \citep{Gaidos_1996,Paliya_2015}. At first sight this is not compatible with our scenario where the size and speed of the perturbation are fitted for about a day to day timescale variability. However we did not consider that these perturbations should naturally be very turbulent environnements. Small scale turbulence crossing a shock is well suited to produce fast flares, as simulated by \cite{Marscher_2014}.

Prior to this study, the likely possibility of multiple high-energy emission zones in the Mrk 421 jet were discussed in many works \citep[e.g.][]{Blazejowski_2005, Balokovic_2016, Carnerero_2017,Kapanadze_2018_2,Kapanadze_2018_1}. While having a general good broadband SED representation, these studies highlighted that the single zone synchrotron self-Compton (SSC) scenario is strongly challenged by some observed variability patterns and also has difficulties to model the hard TeV spectrum.

Due to the high frequency of the synchrotron peak, the SSC interaction falls into the Klein-Nishina regime at TeV energies, preventing any strong radiation \citep{Fossati_2008}. This is a common issue of the so-called "extreme blazars" (EHBLs, or UHBLs), including Mrk 421 \citep{Ghisellini_1999}.
This issue can be resolved if we consider another radiation field in VHE. It encouraged the development of (lepto-) hadronic scenarios, where this additional radiation can be produced by protons (synchrotron or inverse-Compton), or secondary particle emission. Several of these models were addressed to the study of Mrk 421 \citep{Abdo_2011,Mastichiadis_2013,Zech_2017}.

A natural leptonic explanation can however be proposed in the framework of the multiple-shock scenario. If we consider that a small fraction of the particles accelerated in the first shock are not fully cooled before reaching other shocks, they will be re-accelerated. Consecutive shocks have then the potential to push the spectra up to the highest energies in AGN, as shown by \cite{Meli_2013}, and can explain an excess in TeV spectra with respect to one-zone leptonic approach.
It is also interesting to note that this spectral issue mostly occurs in HBLs, which were observed to be the most likely sources to have multiple quasi-stationary knots in their jets \citep{Hervet_2016,Piner_2018}. It then makes HBLs the best candidates for such a particle re-acceleration.

As a last point, we can highlight that variability induced by a change of the thermal and non-thermal particle density crossing a shock (or similar to a shock crossing different density regions) was proposed in various studies of Mrk 421.
From the evolution of Mrk 421 flares, \cite{Fossati_2008} noted that it is "very suggestive of acceleration or injection of the higher energy end of the electron population" as expected by such a multiple-shock re-acceleration process. It was also highlighted  by \cite{Garson_2010} that the variation likely comes  from a change of the local density encountered in the shock environs.
In this view, radiative shock scenarios (whether single, multi-zones, semi-analytic, or MHD based) are promising, such as \cite{Chen_2011,Moraitis_2011, Marscher_2014,Fromm_2016,Bodo_2018}.

\section{Conclusion}

In this paper we show evidence for a possible regular pattern of post-flare variability in Mrk 421.
The time delay of the suggested post-flare excesses in the Mrk 421 stacked lightcurve are consistent with a scenario of the propagation of jets perturbations with roughly similar sizes and constant speeds crossing the multiple stationary VLBI radio knots.

The favored interpretation is a main emission zone in the most upstream VLBI radio knot at 0.38 mas from the core and secondary emission zones from the 3 other downstream radio knots.
This interpretation is preferred at a 3 sigma level to stochastic fluctuations, as reproduced by numerous realistic simulated lightcurves.

From our multiple-flare model fitted to the dataset, we deduce an apparent speed of the flow $\beta_{app} = 45^{+4}_{-2}$  c. It leads to a jet angle with the line of sight $\theta \in [0.38 - 1.8]$ deg, associated with a Lorentz factor $\Gamma \in [43 - 66]$ and a Doppler factor $\delta \geq 31$, and a typical intrinsic size of the perturbations crossing the jet $W_{p,int} \in [0.43 - 19] \times 10^{17}$ cm.
These physical quantities shed a new light on the jet physics of Mrk 421 by providing strong constraints, not based on usual broadband SED models, nor from direct observed motions in jets.

The multiple-shock scenario probed brings a natural and simple solution to the blazar bulk Lorentz factor crisis. Stationary radio knots are interpreted as stationary shocks (likely recollimation shocks), and thus, are not direct markers of the jet flow speed.
The deduced Lorentz and Doppler factor from the multiple-shock scenario are relatively high, but not in disagreement with SSC broadband models and observed fast variability presented in previous studies.
We also note that a very recent study performed by Banerjee et al. (2019, submitted to MNRAS) on a time-dependent modeling of Mrk 421 in internal shock scenario, leads to beaming parameters of Mrk 421 fully consistent with our estimations ($\theta = 1.3$ deg, $\delta \in [40-44]$, $\Gamma \in [28 - 40]$, and $W_{p,int} = 1.1 \times 10^{17}$ cm).
These similar results from a totally independent study and method strengthen the relevance of our approach.

The accuracy of the method can be naturally improved by having long monitoring after strong flares, the larger the dataset, the better an intrinsic post-flare pattern can be distinguished. It would be also improved by a better radio-VLBI monitoring. More radio data will reduce the uncertainties on size and position of radio knots.

This first study probing a post-flare variability pattern in Mrk 421 has considerable potential to be extended in multiple ways. Given the strong X-ray - VHE correlation of Mrk 421, a natural continuity would be to check this pattern in the VHE lightcurves of suitable observatories, such as VERITAS, MAGIC, or FACT.

As soon as a blazar is identified with multiple stationary knots, and has a multi-year dense monitoring in an energy band associated with a great variability (usually in the energy range of its synchrotron or inverse-Compton peaks), it is theoretically possible to perform the same study. 
Confirming such a pattern in multiple other sources would lead to a great leap forward in our knowledge of AGN jet physics and the origin/location of the high energy emission zones.

\acknowledgments

We thank Jonathan Biteau for his helpful suggestions and advises on lightcurve simulation, and Pranati Modumudi for assistance with analysis of the \textit{Swift}-XRT data. We acknowledge support from the U.S. NASA \textit{Swift}-GI grants NNH17ZDA001N and NNX16AN78G. We also thank U.S. National Science Foundation for support under grants PHY-1307311 and PHY-1707432. This research has made use of data from the MOJAVE database that is maintained by the MOJAVE team \citep{Lister_2018}.
\software{XSpec \citep{Arnaud_1996}, astroML \citep{astroML}, Astropy \citep{Astropy_2013}, SciPy \citep{Scipy}, NumPy \citep{Numpy}, Matplotlib \citep{Matplotlib})}

\bibliographystyle{aasjournal}
\bibliography{Var_shock_Mrk421_accepted_arxiv}

\end{document}